\documentclass[
  journal=medium,
  manuscript=article-type,
  year=2020,
  volume=37,
]{cup-journal}

\usepackage{amsmath}
\usepackage[nopatch]{microtype}
\usepackage{booktabs}

\usepackage{amsfonts}       
\usepackage{nicefrac}       
\usepackage{microtype}      
\usepackage{lipsum}
\usepackage{comment}

\usepackage[english]{babel}
\usepackage{graphicx}
\usepackage{multirow}
\usepackage{amsthm}
\usepackage{amssymb}
\usepackage{bbm}
\usepackage[colorinlistoftodos]{todonotes}
\usepackage[colorlinks=true, allcolors=blue]{hyperref}
\usepackage{caption}
\usepackage{float}
\usepackage{subcaption}
\usepackage{multirow}
\usepackage{makecell}
\usepackage{mathtools}
\usepackage{algorithmic}
\usepackage{bm}
\usepackage{physics}
\usepackage{appendix}
\usepackage{enumitem}
\usepackage{csquotes}
\allowdisplaybreaks

\numberwithin{thm}{section}
\numberwithin{defn}{section}
\numberwithin{property}{section}
\numberwithin{lemma}{section}
\numberwithin{cor}{section}

\newcommand{\E}[1]{\ensuremath{\mathbbm{E}\left(#1\right)}}

\newcommand{\bfmu}{\boldsymbol{\mu}}
\newcommand{\bfalpha}{\boldsymbol{\alpha}}
\newcommand{\bfdelta}{\boldsymbol{\delta}}
\newcommand{\bfbeta}{\boldsymbol{\beta}}
\newcommand{\bfy}{\boldsymbol{y}}
\newcommand{\bfOmega}{\boldsymbol{\Omega}}

\title{A Novel Class of Unfolding Models for Binary Preference Data}

\author{Rayleigh Lei}
\affiliation{Department of Statistics, University of Washington, Seattle, 98195, Washington, United States of America}
\email[Rayleigh Lei]{rlei13@uw.edu}

\author{Abel Rodriguez}
\affiliation{Department of Statistics, University of Washington, Seattle, 98195, Washington, United States of America}


\begin{document}

\begin{abstract}
We develop a new class of spatial voting models for binary preference data that can accommodate both monotonic and non-monotonic response functions, and are more flexible than alternative ``unfolding'' models previously introduced in the literature. We then use these models to estimate revealed preferences for legislators in the U.S.\ House of Representatives and justices on the U.S.\ Supreme Court.  The results from these applications indicate that the new models provide superior complexity-adjusted performance to various alternatives and also that the additional flexibility leads to preferences' estimates that more closely match the perceived ideological positions of legislators and justices. 
%
%
%
\end{abstract}

\maketitle

\section{Introduction}

Methods for estimating the preferences of members of deliberative bodies from their voting records have become a foundational tool in political science.  Techniques, such as NOMINATE \citep{poole1985spatial}, IDEAL \citep{JackmanMultidimensionalAnalysisRoll2001}, and its relatives, are used for exploratory purposes to describe the behavior of voters (e.g., see \citealp{jenkins2006impact}, \citealp{poole2006ideology}, \citealp{shor2011ideological} and \citealp{luque2022bayesian}), as well as in confirmatory settings to test specific theories of legislative or judicial behaviour
(e.g., \citealp{schickler2000institutional}, \citealp{schwindt2004gender} and \citealp{clark2012examining}).  However, despite their widespread use, these scaling methods have a number of limitations (e.g., see \citealp{roberts2007statistical}, \citealp{hug2010selection}, \citealp{clinton2012using} and \citealp{jess14-two}).  In particular, in situations where the policy space is assumed to be unidimensional and it is common for voters on both ends of the political spectrum to vote together against those in the middle, standard methods often lead to what could be considered inaccurate estimates for the most extreme legislators (e.g., see \citealp{Lewis2019a,Lewis2019b} and \citealp{yu2021spatial}).  In response to these issues, \citet{duck2022ends} recently introduced a Bayesian version of the generalized graded unfolding model (GGUM, \citealp{roberts2000general}) to the political science literature, and demonstrated that this Bayesian GGUM (BGGUM) is more appropriate than traditional scaling methods for ends-against-middle votes.  

GGUMs originated in the psychology literature, where they have arguably become the most popular ideal point model for non-cognitive measurements that indicate how well an item describes the respondent’s typical preference on a latent continuum.  The key insight behind their construction is that, in one dimensional settings, individuals might disagree with a particular statement when their own preferences lie too far in either direction from a reference point, leading to non-monotonic response functions.  \citet{roberts2000general} describe this phenomenon as individuals potentially disagreeing either ``from above'' and ``from below'', and construct their model for the observed preferences by ``folding'' a traditional latent, ``subjective'' scale with a monotonic response function.

In our view, one shortcoming of the \citet{roberts2000general} and \citet{duck2022ends} approach in the context of political science applications is the lack of a clear, explicit connection with spatial voting models \citep{davis1970expository,enelow1984spatial}.  
A second challenge with the BGGUM is computational.  While various packages implementing algorithms to fit GGUMs exist (e.g, see \citealp{wang2015mcmc}, \citealp{tendeiro2019ggum} and \citealp{tu2021bmggum}), estimation can be challenging in practice.  In particular, in the context of Bayesian inference, many of the methods are based on Metropolis-Hastings algorithms that rely on proposals that need to be carefully calibrated and can fail to properly explore the posterior distribution.  \citet{duck2022ends} presents a Metropolis-coupled MC3 algorithm that uses a set of parallel chains running at various temperatures. This approach shows improved mixing when compared with previous algorithms.  However, their algorithm still relies on a series of random-walk Metropolis-Hastings proposals that need to be carefully calibrated for each new dataset, and our numerical experiments suggest that these calibration can be difficult and time consuming.  Furthermore, the structure of the MC3 algorithm makes extensions of the basic model to hierarchical settings difficult to implement, specially if they involve model comparisons or other discrete random variables (e.g., see \citealp{rodriguez2015measuring}, \citealp{lofland2017assessing} and \citealp{MoserEtAlMultipleIdealPoints2021}).  One final challenge refers to prior specification for BGGUMs.  The most common choice of priors, introduced in \citet{de2006markov} and adopted by \citet{duck2022ends}, involves the use of four-parameter Beta distributions with a compact support.  There is no clear guidance in the literature on how the hyperparameters of these priors should be chosen beyond vague warnings about retrospectively ensuring that the support of the priors contains the support of the corresponding posteriors.

In this paper, we introduce an alternative to the GGUM of \citet{roberts2000general} whose construction relies on the random utility framework of \citet{mcfadden1973conditional}, providing an explicit link between unfolding models and spatial voting models.  We also introduce an efficient Gibbs sampling algorithm that does not require ad-hoc tuning, and discuss criteria for prior specification. The resulting model is used to analyze voting data from the U.S.\ House of Representatives between 1987 and 2022.  In this context, we show that our new unfolding model provides a better complexity-adjusted fit to the data than IDEAL and the BGGUM of \citet{duck2022ends} (as measured by the Watanabe-Akaike Information Criteria, e.g., see \citealp{watanabe2010asymptotic} and \citealp{watanabe2013widely}), as well as estimates of legislators preferences that better fit what would seem to be their true ideological leanings.  We then move to extend this basic framework to create \textit{dynamic} unfolding models in which the latent traits are allowed to evolve over time.  The resulting model, which can be seen as a generalization of the dynamic factor model for binary data introduced by \citet{MartinQuinnDynamicIdealPoint2002a}, is used to analyze voting patterns in U.S.\ Supreme Court between 1935 and 2021.  In this context, we show that voting patterns at key points of the U.S.\ Supreme Court can be better described by our new dynamic unfolding model.

\section{A spatial formulation for unfolding models}\label{sec:staticmodel}

\begin{figure}[!t]
    \centering
    \includegraphics[width = .95\textwidth]{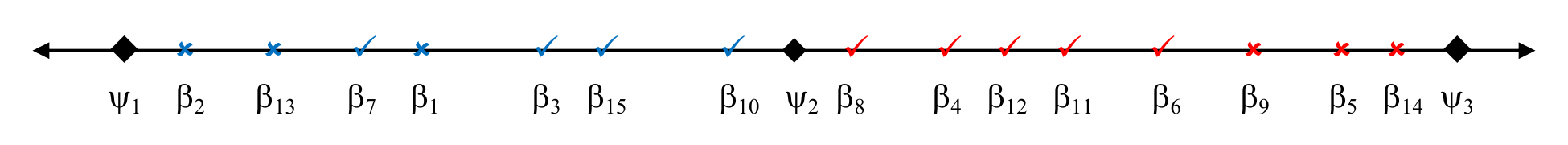}
    \caption{Cartoon representation of the spatial voting model construction of our unfolding model. Squares represent the $\psi_j$'s, check marks represent the ideal points of legislators voting in favor the issue, and crosses represent the ideal points of legislators voting against it.}
    \label{fig:cartoon}
\end{figure}

Let $y_{i, j}$ represent the vote of member $i = 1, 2, \ldots, I$ on issue $j = 1, 2, \ldots, J$. 
We assume that there is a latent one-dimensional Euclidean space (the \textit{policy space}) and that each voter has  a preferred position in that space (their \textit{ideal point}), denoted by $\beta_1, \ldots, \beta_I$.  Additionally, each vote has associated with it three positions, $\psi_{j,1}$, $\psi_{j,2}$ and $\psi_{j,3}$, such that $\psi_{j,2}$ corresponds to the preferred position for a positive (``Aye'') vote on issue $j$, while $\psi_{j,1}$ and $\psi_{j,3}$ are the preferred position for a negative (``Nay'') vote (please see Figure \ref{fig:cartoon}).  In the sequel, we assume that either $\psi_{j,1} < \psi_{j,2} < \psi_{j,3}$ (in which case $\psi_{j,1}$ and $\psi_{j,3}$ correspond to a negative vote because the member disagrees ``from below'' and ``from above'', respectively) or $\psi_{j,3} < \psi_{j,2} < \psi_{j,1}$ (in which $\psi_{j,1}$ and $\psi_{j,3}$ have the opposite interpretation). Under the order constraint, we can interpret one of the ``Nay'' positions (either $\psi_{j,1}$ or $\psi_{j,3}$, depending on the orientation of the latent scale) as the \textit{status quo}, the ``Aye'' position $\psi_{j,2}$ as the proposed policy, and the second ``Nay'' position as an even more extreme policy that some legislators might find preferable to the proposed one.  This interpretation of the latent positions aligns with a common narrative behind extreme-against-the-middle votes in which the most extreme members of the majority party vote against their own party's proposed policy because it is not extreme enough and they are not willing to settle for incremental change. 

Similarly to \citet{JackmanMultidimensionalAnalysisRoll2001}, we assume that members choose among these three options on the basis of quadratic utility functions that depend on the distance between their ideal point $\beta_i$ and the vote positions $\psi_{j,1}$, $\psi_{j,2}$ and $\psi_{j,3}$,
\begin{align*}
    U_{N^{-}}(\beta_i, \psi_{j,1}) &= -\left(\beta_i - \psi_{j,1}\right)^2 + \epsilon_{i,j,1}, \\
    U_{Y}(\beta_i, \psi_{j,2}) &= -\left(\beta_i - \psi_{j,2}\right)^2 + \epsilon_{i,j,2}, \\
    U_{N^{+}}(\beta_i, \psi_{j,1}) &= -\left(\beta_i - \psi_{j,3}\right)^2 + \epsilon_{i,j,3},
\end{align*}
where $\epsilon_{i,j,t,1}$, $\epsilon_{i,j,t,2}$, and $\epsilon_{i,j,3}$ are independent and identically distributed shocks.  Then, an affirmative vote occurs if and only if $U_{Y} > U_{N^{-}}$ and $U_{Y} > U_{N^{+}}$, so that
\begin{align*}
    \textrm{P}(y_{i, j} = 1 \mid \beta_i, \psi_{j,1}, \psi_{j,2}, \psi_{j,3}) &= \textrm{P}(U_{Y}(\beta_i, \psi_{j,2}) > U_{N^{-}}(\beta_i, \psi_{j,2}), U_{Y}(\beta_i, \psi_{j,2}) > U_{N^{+}}(\beta_i, \psi_{j,2}))\\
    &= \textrm{P}(\epsilon_{i,j,1} - \epsilon_{i,j,2} < \alpha_{j,1}(\beta_i - \delta_{j,1}), \epsilon_{i,j,3} - \epsilon_{i,j,2} < \alpha_{j,2}(\beta_i - \delta_{j,2})) ,
\end{align*}
where $\alpha_{j,1} = 2(\psi_{j,2} - \psi_{j,1})$, $\alpha_{j,2} = 2(\psi_{j,2} - \psi_{j,3})$, $\delta_{j,1} = (\psi_{j,1} + \psi_{j,2})/2$ and $\delta_{j,2} = (\psi_{j,3} + \psi_{j,2})/2$.  

In the special case where $\epsilon_{i,j,k}$s are independently and identically distributed from a standard Gumbel distribution, standard theory indicates that 
\begin{align}\label{eq:logisticBGUM}
    \textrm{P}(y_{i, j} = 1 \mid \beta_i, \alpha_{j,1}, \delta_{j,1}, \alpha_{j,2}, \delta_{j,2}) &= \frac{1}{1 + \exp\left\{ \alpha_{j,1}(\beta_i - \delta_{j,1})\right\} + \exp\left\{ \alpha_{j,2}(\beta_i - \delta_{j,2})\right\}}.
\end{align}
Equation \eqref{eq:logisticBGUM} is very similar (although not identical) to that associated with the GGUM for binary data in \citet{roberts2000general} and \citet{duck2022ends}. One key difference is that the traditional GGUM constrains the coefficients $\alpha_{j,1}$ and $\alpha_{j,2}$ to be proportional to each other in a way that seems somewhat arbitrary.  

Alternatively, if the shocks are normally distributed, we obtain a probit unfolding model with response function
\begin{multline}\label{eq:probitunfoldinglik}
    \textrm{P}(y_{i,j} = 1 \mid \beta_i, \alpha_{j,1}, \delta_{j,1}, \alpha_{j,2}, \delta_{j,2}) = \\
    \int_{-\infty}^{\alpha_{j,1}(\beta_i - \delta_{j,1})} \int_{-\infty}^{\alpha_{j,2}(\beta_i - \delta_{j,2})} \frac{1}{2 \sqrt{3} \pi} \exp\left\{ -\frac{1}{3} (z_1^2 - z_1 z_2 + z_2^2) \right\} \dd z_1 \dd z_2.
\end{multline}
This is simply the cumulative distribution function of the bivariate normal distribution with mean $\mathbf{0}$, and covariance matrix 
$
\begin{pmatrix}
2 & 1    \\
1 & 2
\end{pmatrix}
$ evaluated at $(\alpha_{j,1}(\beta_i - \delta_{j,1}), \alpha_{j,2}(\beta_i - \delta_{j,2}))$. In the remainder of this paper we will focus on this probit version of the unfolding model.

An appealing property of our construction is that it clearly includes IDEAL as a special case.  For example, under the assumption of Gaussian shocks, and if we have $\psi_{j,1} < \psi_{j,2} < \psi_{j,3}$, letting $\psi_{j,3} \to \infty$ implies that
\begin{align}\label{eq:assymptoticIDEAL}
    \textrm{P}(y_{i, j} = 1 \mid \beta_i, \alpha_{j,1}, \delta_{j,1}, \alpha_{j,2}, \delta_{j,2}) &\to \Phi\left( \alpha_{j,1}(\beta_i - \delta_{j,1}) \right) ,
\end{align}
where $\Phi(\cdot)$ denotes the cumulative distribution function of the standard normal distribution.

\subsection{Prior distributions}\label{sec:priors}

We adopt a Bayesian approach to inference and proceed to discuss the prior distributions associated with the unknown model parameters $\beta_1, \ldots, \beta_I$, $\bfalpha_1, \ldots, \bfalpha_J$ and $\bfdelta_1, \ldots, \bfdelta_J$.  We first discuss the functional form of the priors, which is chosen in part for computational convenience, and then discuss hyperparameter selection.

The choice for the prior distribution on the ideal points $\beta_1, \ldots, \beta_I$ is straightforward.  We follow most of the literature and let the ideal points be independent and identically distributed from a standard normal distribution, $\beta_i \sim \textrm{N} \left( 0, 1 \right)$.  Fixing the mean and variance of this distribution helps in addressing some of the identifiability issues associated with spatial voting models (see Section \ref{sec:identifiability}).

The design of priors for the vote-specific parameters $\bfalpha_{j} = (\alpha_{j,1}, \alpha_{j,1})'$ and $\bfdelta_{j} =(\delta_{j,1}, \delta_{j,2})'$ is more difficult.  One consideration to keep in mind  is that there is no natural ``correct'' direction for the underlying policy space, so it is important that $\alpha_{j,1}$ and $\alpha_{j,2}$ marginally be allowed to have full support over the whole real line.  A second consideration is that our original construction requires either $\psi_{j,1} < \psi_{j,2} < \psi_{j,3}$ or $\psi_{j,3} < \psi_{j,2} < \psi_{j,1}$. In addition to facilitating the interpretation of the model, there are at least two more reasons why introducing this constraint is important.  The first one is identifiability.  Indeed, note that if $\psi_{j,1}$ and $\psi_{j,3}$ (the ``Nay'' positions) are on the ``same side'' of $\psi_{j,2}$ (the ``Aye'' position), there is no way to separately learn both $\psi_{j,1}$ and $\psi_{j,3}$ from the data since they are observationally equivalent. Connected to this, we note that the response function is automatically monotonic when $\psi_{j,1}$ and $\psi_{j,3}$ are on the same side of $\psi_{j,2}$.  Hence, without the order constraint, there would be two ways to represent ``traditional'' partisan votes (which correspond to monotonic response functions): (a) by placing both $\psi_{j,1}$ and $\psi_{j,3}$ on the same side of $\psi_{j,2}$, or (b) by putting them on opposite sides of $\psi_{j,2}$ but making one of them very extreme (recall Equation \eqref{eq:assymptoticIDEAL}).  This would lead to a multimodal posterior distribution, which would be challenging to explore.  Hence, our priors force $\alpha_{j,1}$ and $\alpha_{j,2}$ to have opposite signs, which is a sufficient condition to ensure that $\psi_{j,1} < \psi_{j,2} < \psi_{j,3}$ or $\psi_{j,3} < \psi_{j,2} < \psi_{j,1}$.

Based on these considerations, we assign $\bfalpha_{j}$ and $\bfdelta_{j}$ a joint prior that is a mixture of two truncated multivariate Gaussian distributions with density
\begin{multline}\label{eq:alphadeltaprior}
p \left( \bfalpha_{j}, \bfdelta_{j} \right)
      =\frac{1}{32 \pi^2 \omega^2 \kappa^2} \exp\left\{ -\frac{1}{2} \left(\frac{1}{\omega^2 }\bfalpha_j'\bfalpha_j  + \frac{1}{\kappa^2}(\bfdelta_j - \bfmu)'(\bfdelta_j - \bfmu)\right)\right\} \mathbf{1}(\alpha_{j,1}>0, \alpha_{j,2}<0) \\
      + \frac{1}{32 \pi^2 \omega^2 \kappa^2} \exp\left\{ -\frac{1}{2} \left(\frac{1}{\omega^2 }\bfalpha_j'\bfalpha_j  + \frac{1}{\kappa^2}(\bfdelta_j + \bfmu)'(\bfdelta_j + \bfmu)\right)\right\} \mathbf{1}(\alpha_{j,1}<0, \alpha_{j,2}>0) .
\end{multline}

Under this prior, the 
joint density of $\bfalpha$ satisfies the required sign constraints, but the marginal distributions of both $\alpha_{j,1}$ and $\alpha_{j,2}$ correspond to zero-mean univariate Gaussian distributions with variance $\omega^2$ and, therefore have full support over the real line. Furthermore, note that this prior satisfies $p(\bfalpha_{j},\bfdelta_{j}) = p(-\bfalpha_{j},-\bfdelta_{j})$, making the prior invariant to reflections of the latent space.  This property enables us to address identifiability issues related to reflections of the latent space as a post-processing step (please see Section \ref{sec:identifiability}).

To select the hyperparameters $\bfmu$, $\omega$ and $\kappa$, we focus our attention on the implied prior on the parameter $\theta_{i,j} = \mathrm{P}(y_{i,j} = 1 \mid \beta_i, \alpha_{j,1}, \delta_{j,1}, \alpha_{j,2}, \delta_{j,2})$, which is interpretable and comparable across various model formulations.  Here, we again follow the literature and target a prior distribution for $\theta_{i,j}$ that is bimodal, placing most of its probability around $0$ and $1$ (e.g., see \citealp{spirling2010identifying}, \citealp{yu2021spatial} and \citealp{paganin2021computational}).  In practical terms, this assumption means that most of the time, voters are fairly certain about whether they support or oppose a particular measure.  Furthermore, because most measures tend to pass, we expect the distribution of $\theta_i$ to place slightly more probability on values close to $1$.  While there are many priors on $\bfmu$, $\omega$ and $\kappa$ that would satisfy these requirements, we set $\bfmu = (-2,10)'$, $\omega^2 = 25$ and $\kappa^2 = 10$ in the illustrations we discuss in this paper and study the sensitivity of the results to moderate changes in the hyperparameters. Please see Section 4 of the Supplementary Materials.  Figure \ref{fig:prior_draws} presents a histogram for $\theta_{i,j}$ based on $10,000$ draws from our chosen prior distribution, and compares it again a second histogram for the same parameter under the model used in \citet{duck2022ends}.  It is worthwhile noting that the prior for $\theta_{i,j}$ under both model is very similar, enabling some of the comparisons we present in Section \ref{sec:resultsUSHouse}.

\begin{figure}[!t]
\centering
\begin{subfigure}[t]{.45\textwidth}
    \centering
    \includegraphics[width = .95\textwidth]{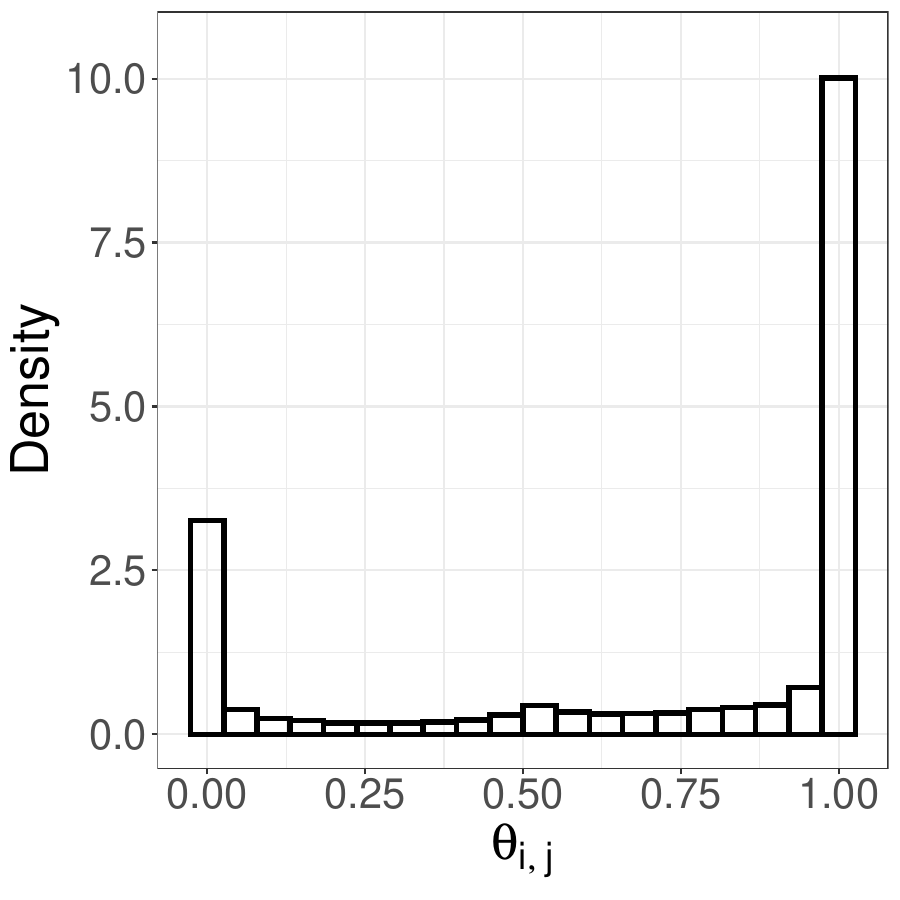}
    \caption{Under our probit unfolding model}
\end{subfigure}
\begin{subfigure}[t]{.45\textwidth}
    \centering
    \includegraphics[width = .95\textwidth]{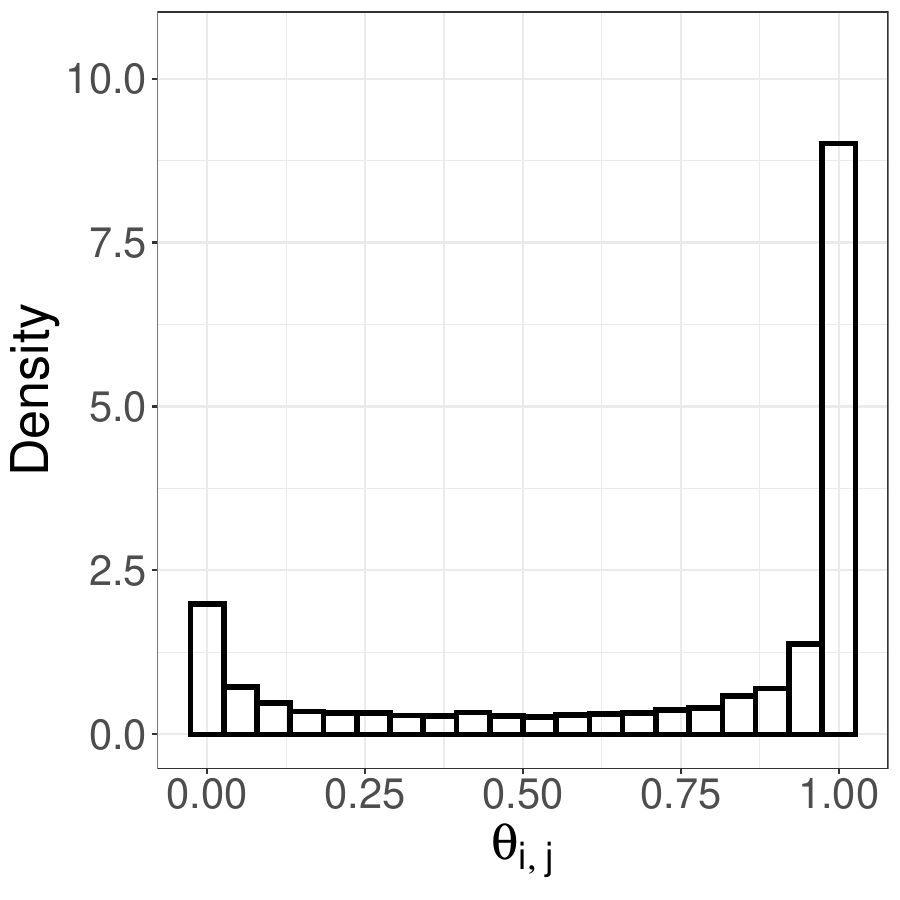}
    \caption{Under the GGUM of \citet{duck2022ends}}
\end{subfigure}
\caption{Histograms for 10,000 draws of the implied prior distribution on $\theta_{i,j}$ for our probit GGUM model under a prior with $\bfmu = (-2,10)'$, $\omega^2 = 25$ and $\kappa^2 = 10$ compared against the implied prior for the same parameter under the model used in \citet{duck2022ends}.}
\label{fig:prior_draws}
\end{figure}

\subsection{Identifiability}\label{sec:identifiability}

Like other spatial voting models, the parameters associated with the latent space are identified in the likelihood only up to an affine transformation.  However, the choice of a standard Gaussian prior for the ideal points $\beta_1, \ldots, \beta_I$ means that the posterior is invariant to shifts and rescalings of the latent space, and therefore, weakly identifiable.  On the other hand, we address invariance to reflections by fixing the sign of the ideal point of one particular legislator.  For example, in the case of the U.S.\ House of Representatives, we fix the sign of the ideal point of the whip of the Republican party to be positive.

\subsection{Computation}\label{sec:computation}

We explore the posterior distribution  using a Markov chain Monte Carlo algorithm that relies on a data augmentation approach similar to that introduced in \citet{albert1993bayesian}. In particular, for each observations $y_{i,j}$ we introduce three auxiliary variables, $y^{*}_{i,j,1}$, $y^{*}_{i,j,2}$ and $y^{*}_{i,j,3}$, which follow a joint multivariate normal distribution on the form
\begin{align}\label{eq:auxrepresentation}
    \begin{pmatrix}
        y^*_{i, j, 1}\\
        y^*_{i, j, 2}\\
        y^*_{i, j, 3}\\
    \end{pmatrix}
    \Bigg |
    \alpha_{j,1}, \alpha_{j,2}, \delta_{j,1}, \delta_{j,2}, \beta_i
    &\sim
    \textrm{N}\left(\begin{pmatrix}
        -\alpha_{j,1}(\beta_i - \delta_{j,1})\\
        0\\
        -\alpha_{j,2}(\beta_i - \delta_{j,2})\\
    \end{pmatrix} ,
    \begin{pmatrix}
        1 & 0 & 0\\
        0 & 1 & 0\\
        0 & 0 & 1\\
    \end{pmatrix}\right) .
\end{align}

Conditional on these auxiliary variables, the unknown model parameters can be sampled from (mixtures of truncated) Gaussian distributions.  Similarly, conditional on the parameters of interest, the full conditional distributions for the auxiliary variables are truncated Gaussians, with the truncation region being determined by the value of $y_{i,j}$.  Hence, it is possible to directly sample from all full conditional posterior distributions and there is no need to tune proposal distributions to specific datasets.  Further details of the algorithm are presented in Section 1 of the Supplementary Materials, and code implementing the algorithm is available at \url{https://github.com/rayleigh/probit_unfolding_model/}.

\section{Revealed preferences in the U.S.\ House of Representative, 1987--2022}\label{sec:resultsUSHouse}

We illustrate the performance of our probit unfolding model by analyzing roll-call voting data from the 100\textsuperscript{th} to the 117\textsuperscript{th} U.S.\ House of Representatives.  We exclude from the analysis legislators who were absent for more than 40\% of the vote, as well as all unanimous votes.  Then, we treat any remaining missing votes as missing completely at random. 

We compare our model against IDEAL \citep{JackmanMultidimensionalAnalysisRoll2001}, as well as the BGGUM of \citet{duck2022ends}.  Posterior summaries for our model are based on 20,000 iterations of our algorithm obtained after burning the first 200,000 samples and thinning the next 200,000 by a tenth.  For IDEAL, we use the algorithm implemented in the \texttt{R} package \texttt{MCMCpack}, and inferences are based on 20,000 iterations obtained after burning the first 10,000 samples.  Finally, computation for the BGGUM relies on the algorithm implemented in the \texttt{R} package \texttt{bggum}, and inferences are based on 20,000 samples obtained after burning the first 5,000 iterations and using the next 5,000 to tune the proposal distributions. Please see \citet{duck2022ends} for details on the \texttt{bggum} package.  Convergence of the various algorithms was checked by monitoring the (unnormalized) joint posterior distribution as well as the ideal points of a few legislators in each House using the multi-chain method of \citet{GelmanRubinInferenceIterativeSimulation1992}.
\begin{figure}[!t]
\centering
    \centering
    \begin{subfigure}[t]{.49\textwidth}
    \includegraphics[width = \textwidth]{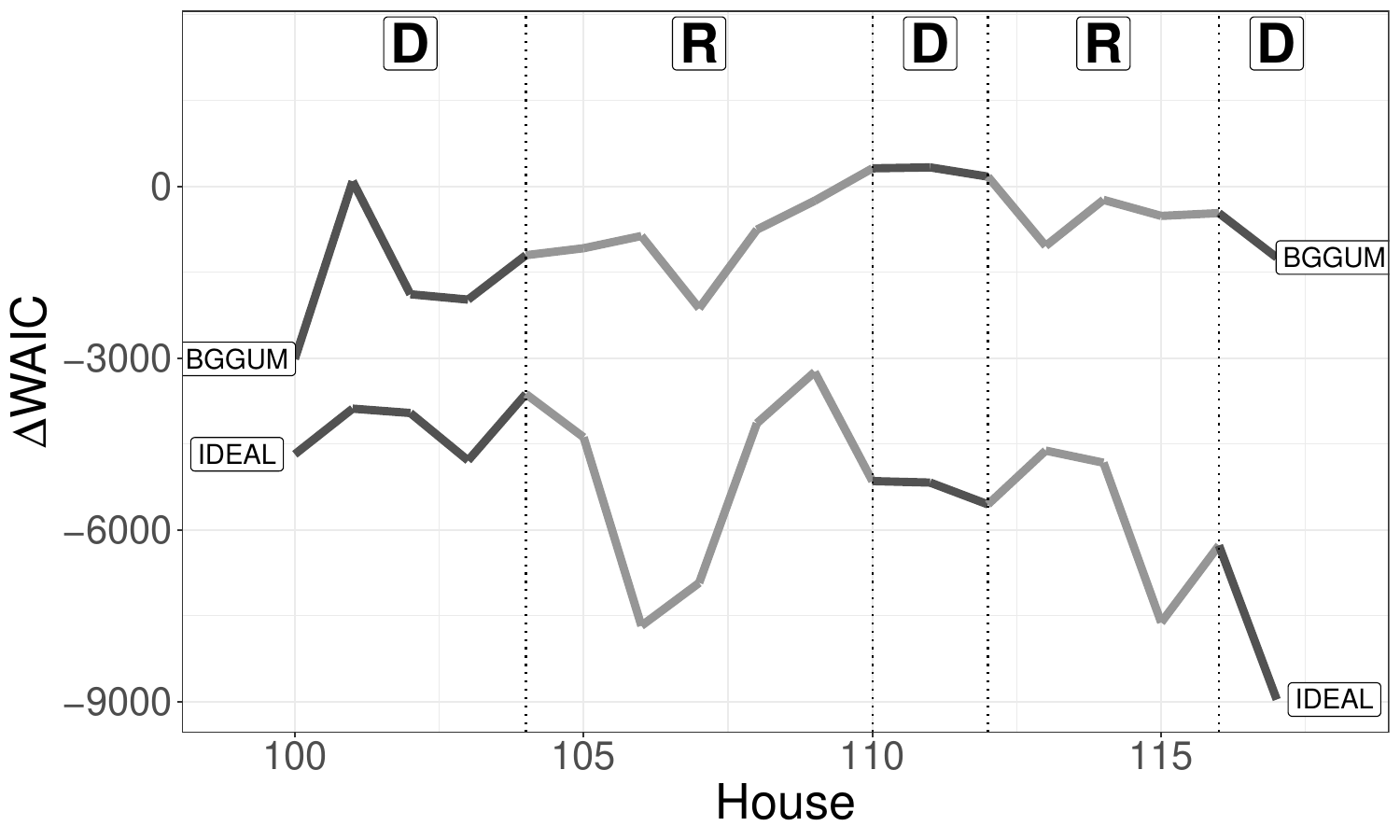}
    \caption{WAIC relative to PUM}\label{fig:static_waic_comp}
    \end{subfigure}
    \begin{subfigure}[t]{.49\textwidth}
    \includegraphics[width = \textwidth]{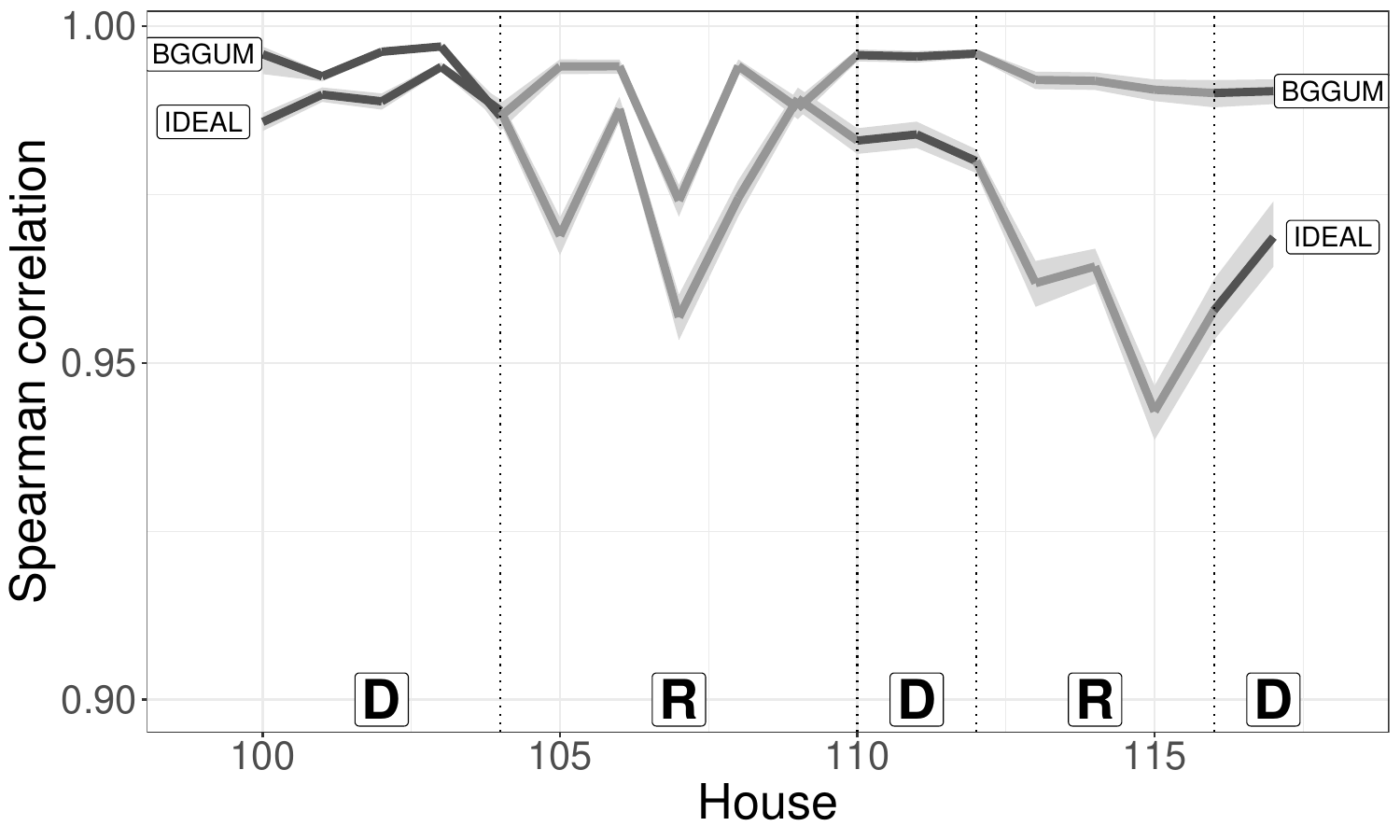}
    \caption{Spearman Correlation relative to the probit unfolding model}\label{fig:static_spearman}
    \end{subfigure}
\caption{Left panel:  Difference in WAIC scores between the probit unfolding model and IDEAL,  ($WAIC(\text{PUM}) - WAIC(\text{IDEAL})$), and between the probit unfolding model and BGGUM,  ($WAIC(\text{PUM}) - WAIC(\text{BGGUM})$). Right panel:  Spearman correlation between the legislator's rankings generated by the probit unfolding model and those generated by either IDEAL or BGGUM.}
\end{figure}

We start by comparing the fit of the various models using a blocked version of the Watanabe-Akaike Information Criterion (WAIC,  \citealp{watanabe2010asymptotic, watanabe2013widely, gelman2014understanding}).  For a given House, the  WAIC for model $m$ is given by
\begin{multline}\label{eq:lWAIC}
    WAIC(m) = -2\left[\sum_{i=1}^{I} \log\left( \textrm{E}_{\textrm{post}} \left\{ 
    \prod_{j = 1}^J \theta_{i,j}(m)^{y_{i,j}} \left[1 -\theta_{i,j}(m)\right]^{1 - y_{i,j}} \right\} \right) \right. \\
     \left. - \sum_{i=1}^{I}  \textrm{var}_{\textrm{post}}\left\{ 
    \sum_{j = 1}^{J} \left[ y_{i,j} \log \theta_{i,j}(m) + (1-y_{i,j}) \log(1-\theta_{i,j}(m)) \right] \right\} \right] ,
\end{multline}
where $\theta_{i, j}(m)$ represents the probability that legislator $i$ votes ``Aye'' in issue $j$ under model $m$. Lower values of the WAIC provide evidence in favor of that particular model. Like the Akaike Information Criterion and Bayesian Information Criterion, the WAIC attempts to balance model fit with model complexity. However, unlike these two criteria, the WAIC is appropriate for hierarchical setting and is invariant to reparametrizations of the model.
\begin{figure}[!t]
\centering
\begin{subfigure}[t]{0.40\textwidth}
    \centering
    \includegraphics[width = \textwidth]{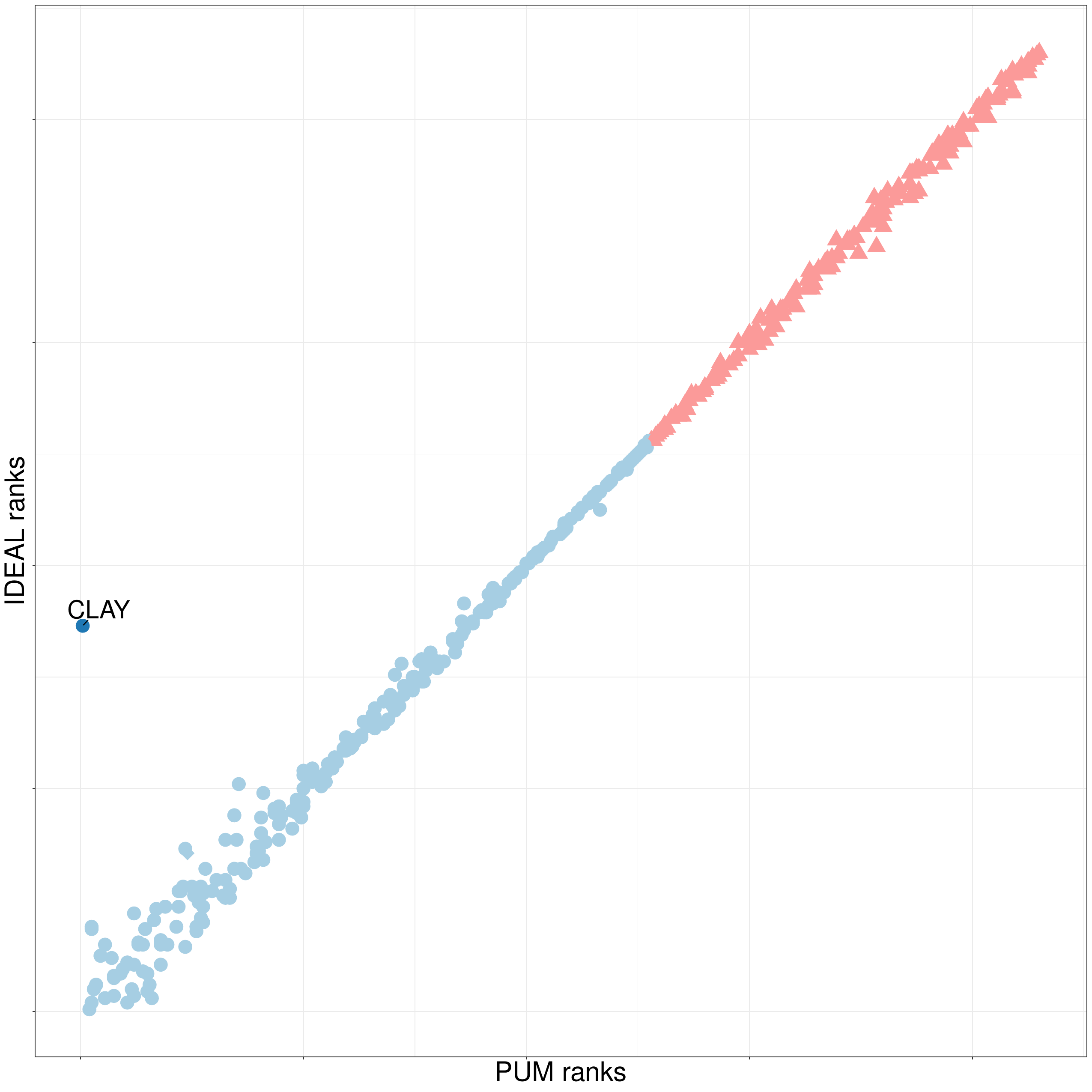}
    \caption{103\textsuperscript{rd} House: Probit unfolding model vs.\ IDEAL}
\end{subfigure} \hspace{12mm}
\begin{subfigure}[t]{0.40\textwidth}
    \centering
    \includegraphics[width = \textwidth]{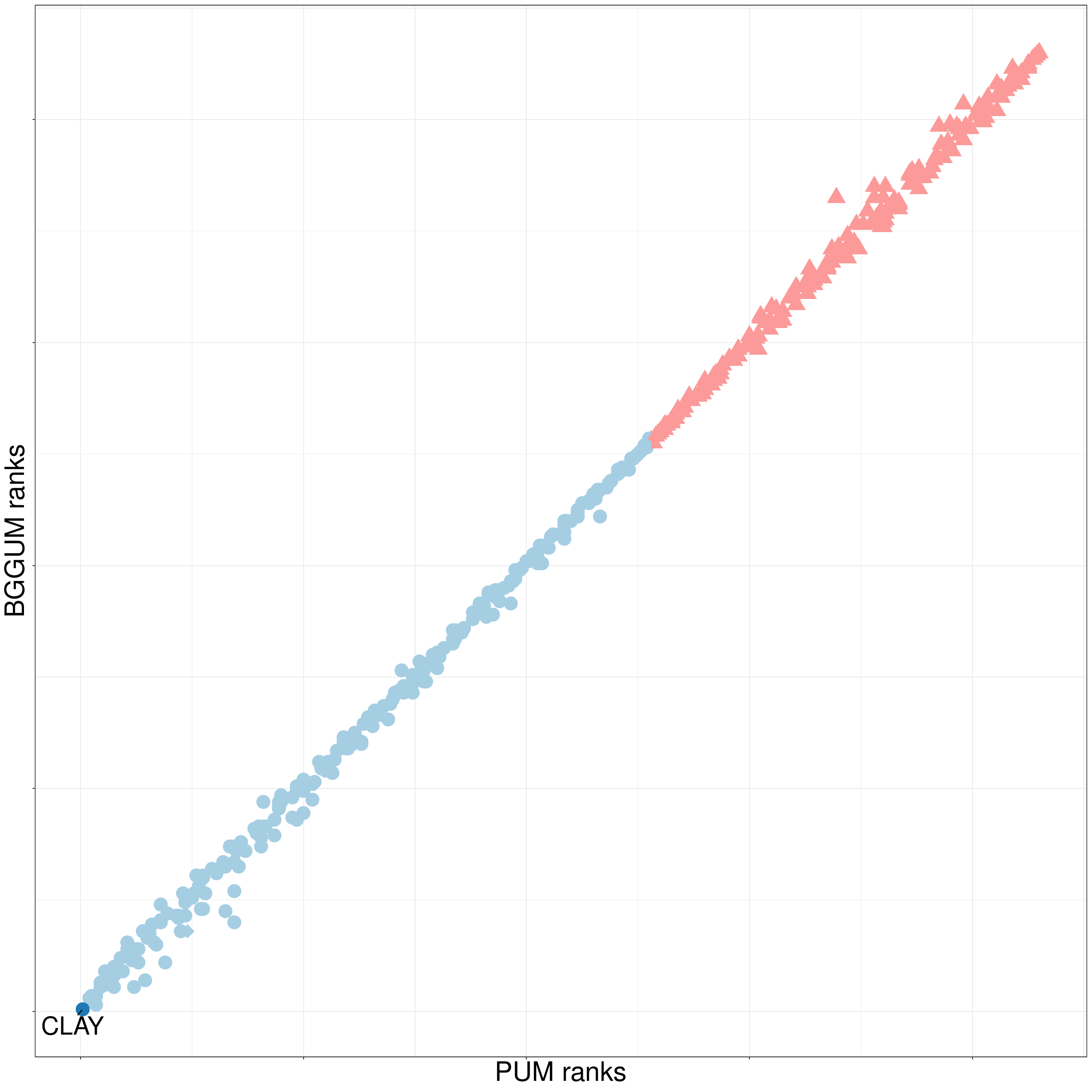}
    \caption{103\textsuperscript{rd} House: Probit unfolding model vs.\ BGGUM}
\end{subfigure}\\
\begin{subfigure}[t]{0.40\textwidth}
    \centering
    \includegraphics[width = \textwidth]{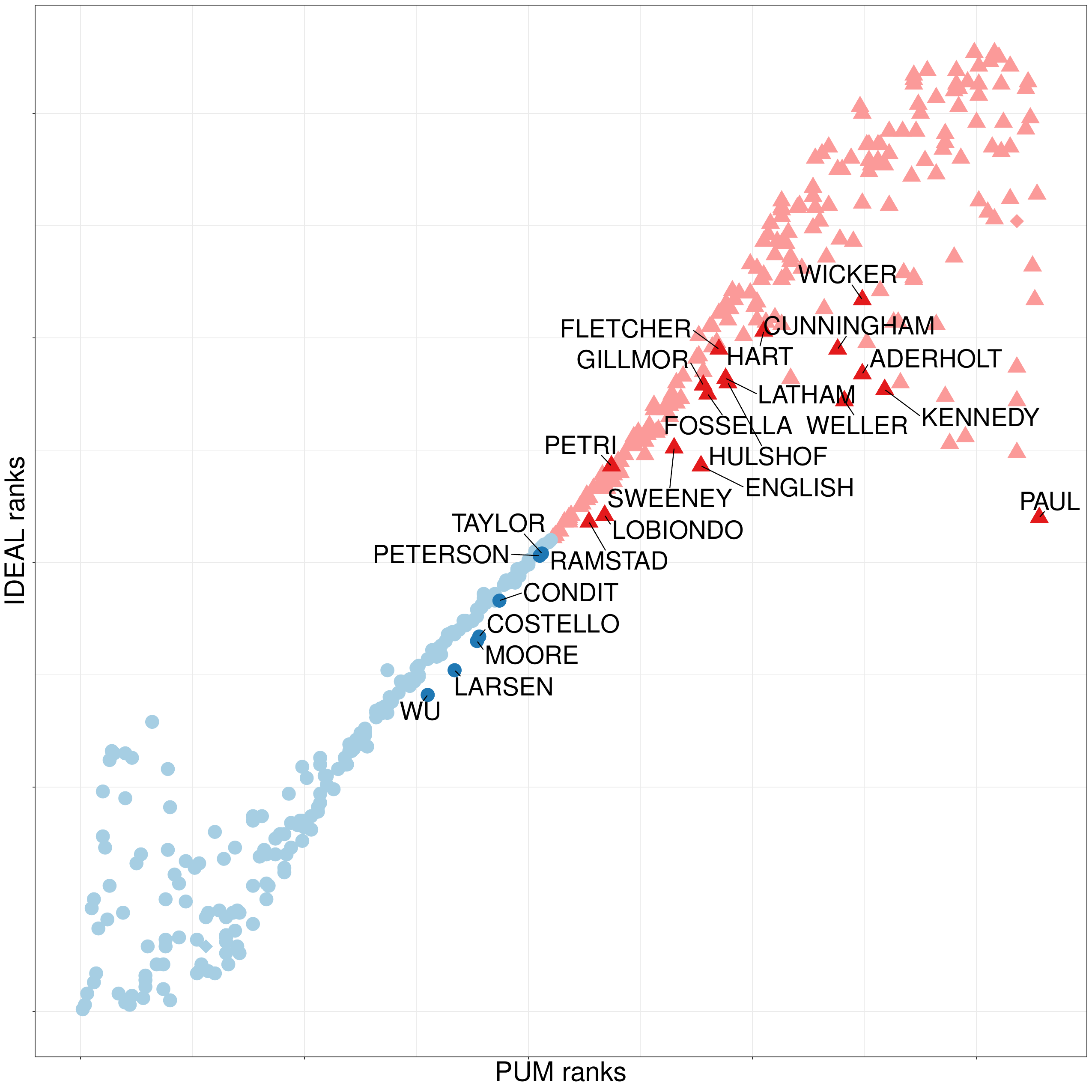}
    \caption{107\textsuperscript{th} House: Probit unfolding model vs.\ IDEAL}
\end{subfigure} \hspace{12mm}
\begin{subfigure}[t]{0.40\textwidth}
    \centering
    \includegraphics[width = \textwidth]{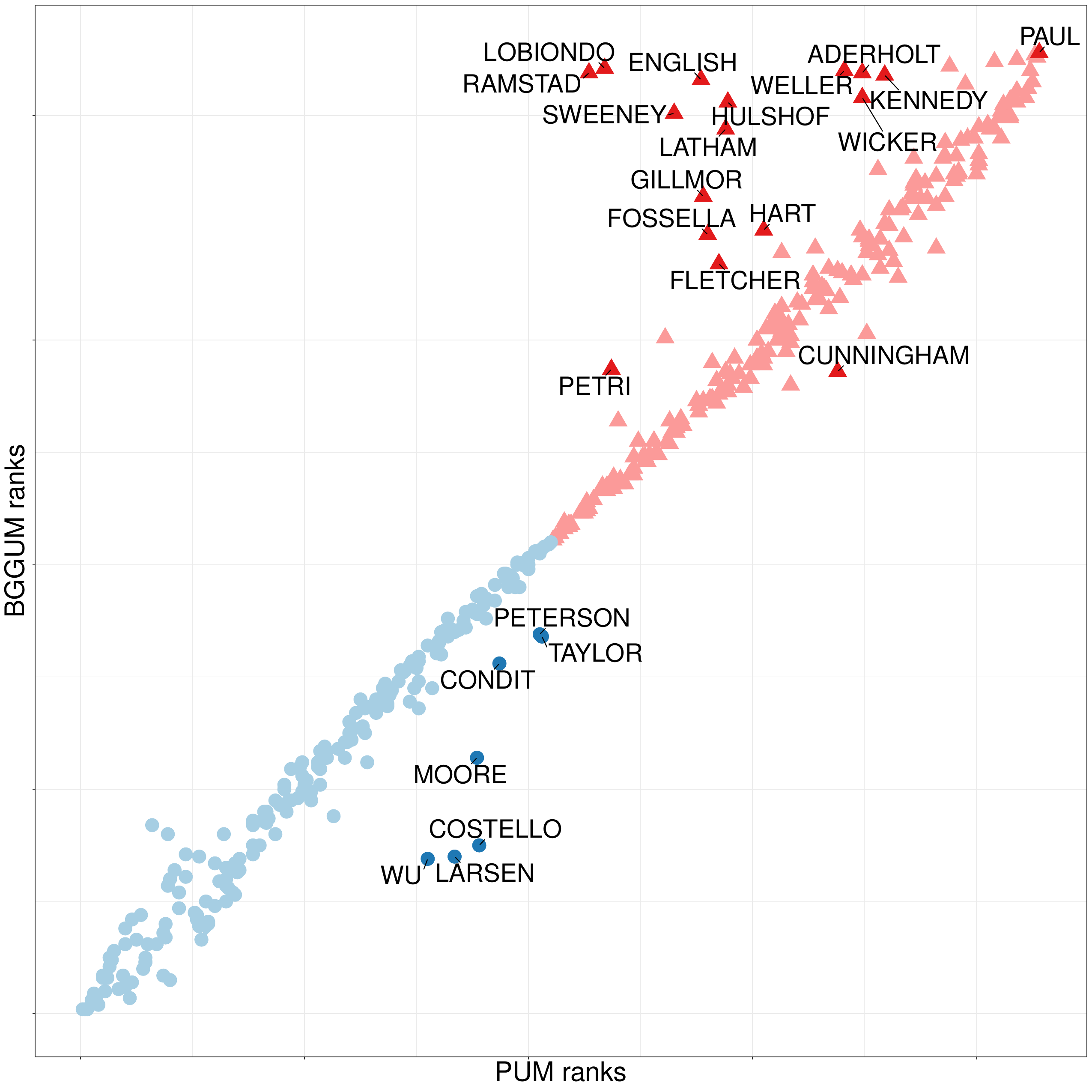}
    \caption{107\textsuperscript{th} House: Probit unfolding model vs.\ BGGUM}\label{fig:rank_comparison_static_107_BBUMPUM}
\end{subfigure}\\
\begin{subfigure}[t]{0.40\textwidth}
    \centering
    \includegraphics[width = \textwidth]{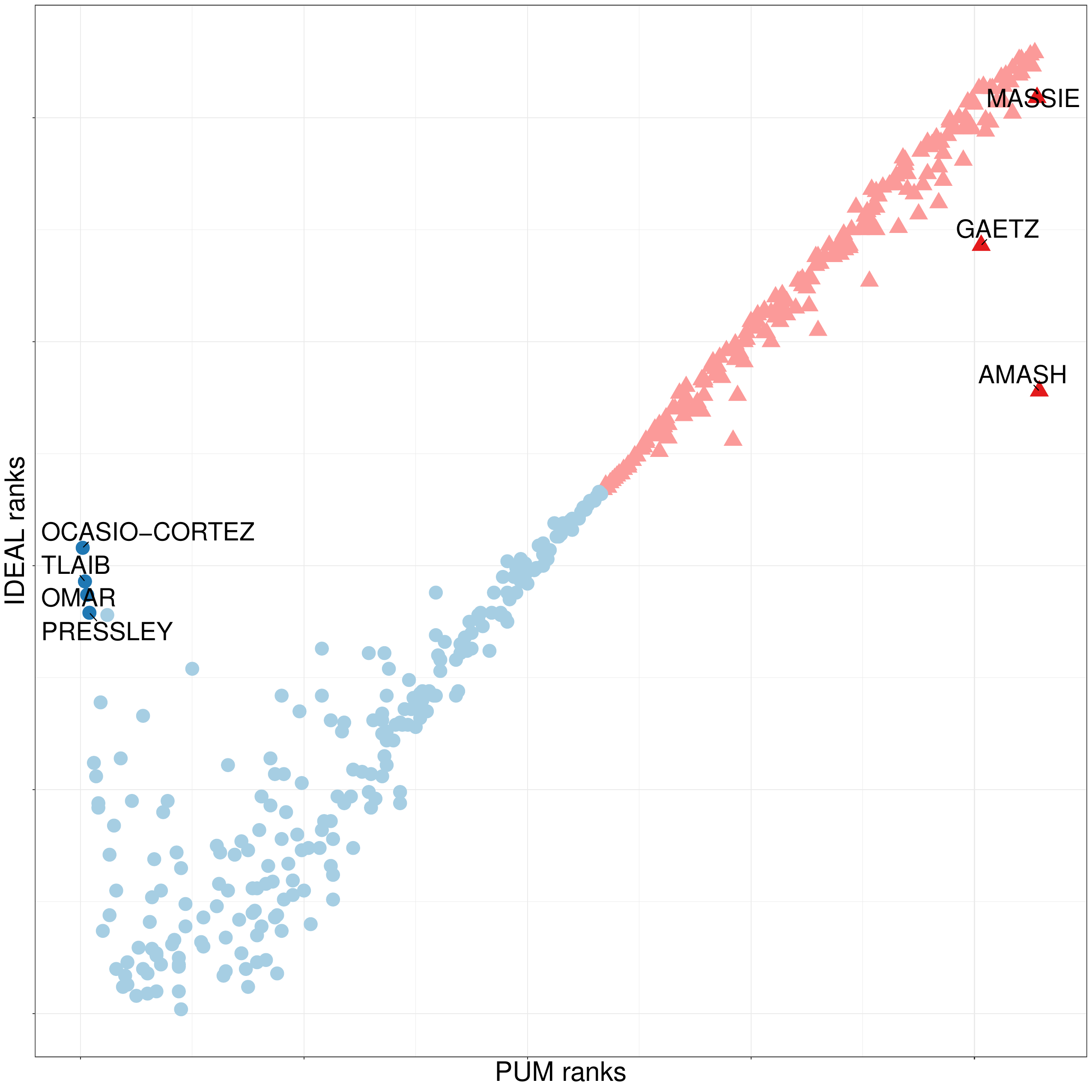}
    \caption{116\textsuperscript{th} House: Probit unfolding model vs.\ IDEAL}
\end{subfigure} \hspace{12mm}
\begin{subfigure}[t]{0.40\textwidth}
    \centering
    \includegraphics[width = \textwidth]{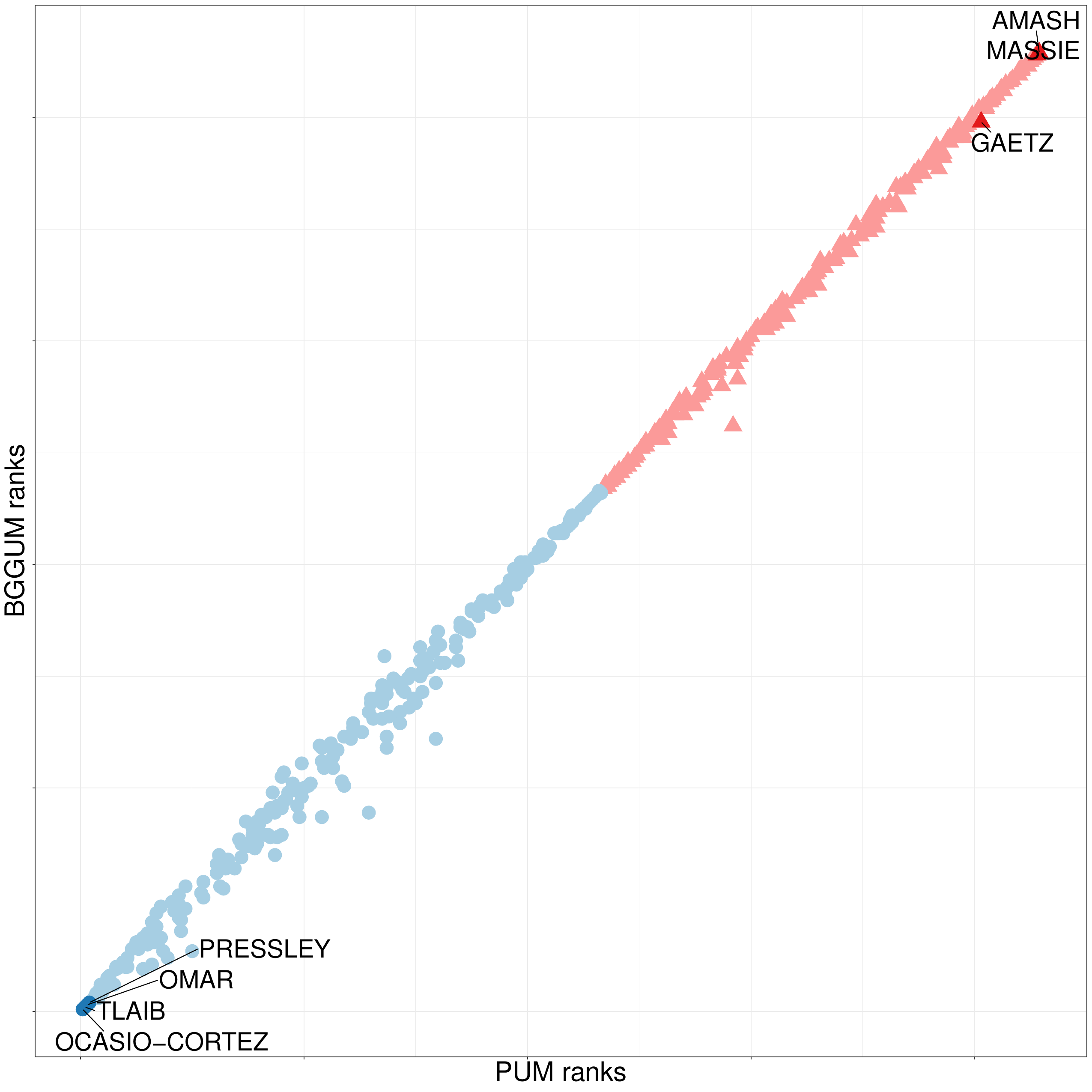}
    \caption{116\textsuperscript{th} House: Probit unfolding model vs.\ BGGUM}
\end{subfigure}
\caption{Comparison of the posterior median ranks of legislators across IDEAL, BGGUM and our probit unfolding model in selected Houses. Democrats are shown with blue triangles, Republicans are shown with red triangles, and independents are shown with a rhombus and the color of the party that they caucus with.}
\label{fig:rank_comparison_static}
\end{figure}

Figure \ref{fig:static_waic_comp} presents 
the difference in WAIC scores between the probit unfolding model and IDEAL, and between the probit unfolding model and BGGUM.  Note that the curve for IDEAL is negative, which is strong evidence that the probit unfolding model outperforms it.  Furthermore, we can see that our probit unfolding model outperforms BGGUM in all but four Houses: the 101\textsuperscript{st}, the 110\textsuperscript{th}, the
111\textsuperscript{th}, and the 112\textsuperscript{th}.  Even in these four cases, the advantage of BGGUM is very small.  
On the other hand, Figure \ref{fig:static_spearman} presents the posterior mean (solid lines) and corresponding 95\% credible intervals (shaded region) for the Spearman correlation between the legislators’ rankings generated by the probit unfolding model and those generated by IDEAL and BGGUM.  The Spearman correlation (which, as the name suggests, lies in the $[-1,1]$ interval) is widely used as a distance between rankings, with values close to 1 indicating that the ranking of the legislators generated by both models are identical \citep{kumar2010generalized}.  As would be expected, the rankings  based on our probit unfolding model are more similar to those generated by BGGUM than those generated by IDEAL.  However, the rankings of the two unfolding models do seem to differ in important ways between the 104\textsuperscript{th} and the 107\textsuperscript{th} Houses.  On the other hand, starting with the 110\textsuperscript{th} House, we see that rankings of the two unfolding models are very close, but tend to differ substantially from the rankings generated by IDEAL.

To gain additional insight into the behavior of the models, we investigate in more detail the estimates of the rankings for the 103\textsuperscript{rd}, 107\textsuperscript{th}, and 116\textsuperscript{th} Houses.  The graphs on the left column of Figure \ref{fig:rank_comparison_static} compare the posterior median ranks generated by the probit unfolding model against those generated by IDEAL, while the right column compares the probit unfolding model against BGGUM.  In line with the results from Figure \ref{fig:static_spearman}, we see that all three models tend to yield very similar rankings during the 103\textsuperscript{rd} House.  The main outlier is representative William Lacy Clay Jr.\ of Missouri. In this case, both unfolding models categorize Clay as more liberal than IDEAL.  On the other hand, for the 116\textsuperscript{th} House, the rankings generated by the two unfolding models are very similar, but quite different from those generated by IDEAL.  In particular, there are large differences on the rankings of many Democratic legislators, specifically those of the so-called ``Squad'' (Representatives Alexandria Ocasio-Cortez of New York, Ilhan Omar of Minnesota, Ayanna Pressley of Massachusetts, and Rashida Tlaib of Michigan).  While both unfolding models rank the members of the Squad as being among the most liberal in the Democratic caucus, IDEAL ranks them as centrists.  Based on what we know about the Squad, the rankings generated by the unfolding models seem much more appropriate.  The members of the Squad were all elected for the first time in 2018 with support by the Justice Democrats political action committee, and are widely known for advocating  progressive policies such as Medicare for All, the Green New Deal, and full student loan debt cancellation.  IDEAL tends to rank them as centrists during the 116\textsuperscript{th} House because they often clashed with the Democratic leadership and selectively voted against their own party, often in conjunction with Republican legislators largely understood to be on the extreme of their own party \citep{Lewis2019b,Lewis2019a}.  There are also some important differences in the rankings of some Republican legislators during the 116\textsuperscript{th} House, particularly Justin Amash of Michigan and Matt Gaetz of Florida, which are ranked as extremes by the unfolding models.  Similar to the Squad on the Democratic side, these legislators are well known members of the more extreme wing of the Republican party and often vote with extreme Democrats and against their party on a number of key issues.  

The previous results suggest that revealed preferences estimated by unfolding models are generally better than those from IDEAL at capturing the ideological leanings of legislators, especially for  more extreme ones.  However, they do not provide much insight into the differences between the two unfolding models.  To address this question, we focus now on the 107\textsuperscript{th} House, which is the one for which Figure \ref{fig:static_spearman} shows the largest difference in rankings between the probit unfolding model and BGGUM.  In the case of Democratic legislators, both unfolding models mostly agree with each other and produce rankings for the most extreme Democrats that are quite different from IDEAL.  Nonetheless, the probit unfolding model and BGGUM yield relative large differences in the rankings of Rick Larsen from Washington, David Wu from Oregon, Jerry Costello from Illinois, Dennis Moore from Kansas, Gary Condit from California, Collin Peterson from Minnesota, and Gene Taylor from Mississippi. In all seven case, the probit unfolding model ranks these legislators as more centrist than BGGUM.  Reviewing their records, it is worthwhile noting that Larsen, Wu and Moore were all members of the centrist New Democrat Coalition, while Condit, Peterson and Taylor were all members of the also centrist Blue Dog Coalition (co-founded by Peterson). This suggests that the rankings generated by the probit unfolding model are more appropriate than those of BGGUM in this case.  On the other hand, Figure \ref{fig:rank_comparison_static_107_BBUMPUM} highlights, among others, the 15 Republican legislators with the largest differences in median ranks.  Of these, Jim Ramstad from Minnesota, Frank Lobiondo from New Jersey, Phil English from Pennsylvania, Paul Gillmor from Ohio, Jerry Weller from Illinois, Tom Petri from Wisconsin and Melissa Hart from Pennsylvania were all members of the moderate Republican Main Street Partnership and consistently ranked highly on bipartisanship indexes.  Similarly, John Sweeney from New York, Kenny Hulshof from Missouri, Vito Fossella from New York and Mark Kennedy from Minnesota had relatively low rankings among Republicans (signaling more moderate ideologies) in Govtrack scores \citep{govtrack}.  On the other hand, Duke Cunningham (R CA-51), the only one in the list whose rank under the probit unfolding model is more conservative than under BGGUM, was widely considered an ardent conservative. Again, this suggests that the rankings produced by the probit unfolding model tend to be a more accurate reflection of legislator's ideology.

%
\begin{figure}[!t]
\centering
    \includegraphics[width = 0.75\textwidth]{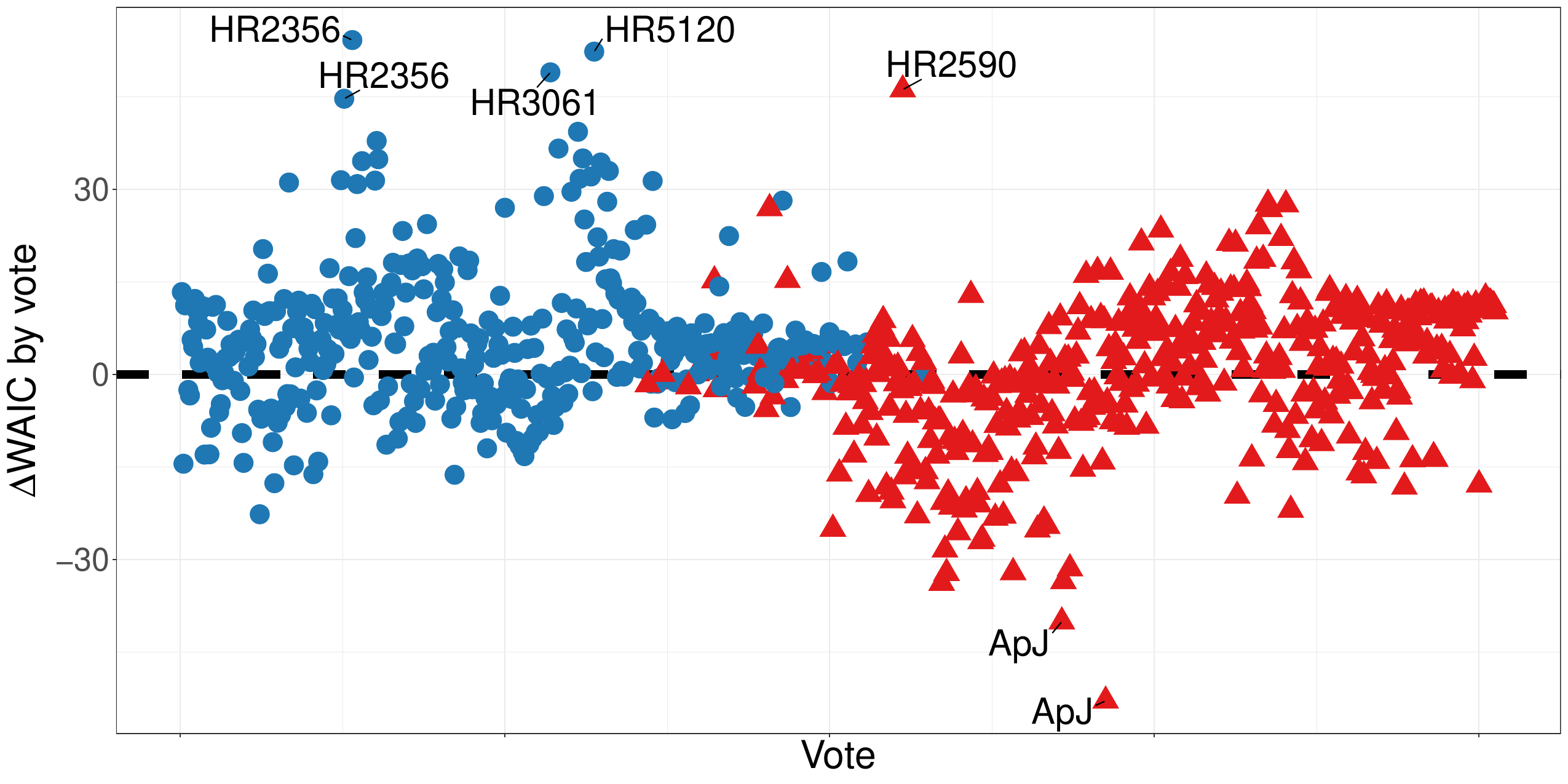}
\caption{Difference in vote-specific WAIC scores between the probit unfolding model and BGGUM, (WAIC(BGGUM) – WAIC(PUM)), for the 107\textsuperscript{th} House. Note that the way the difference is being computed here is the opposite to the way in which it was computed in Figure \ref{fig:static_waic_comp}. Votes are ordered according to the absolute difference between Republican votes and Democratic votes. Blue circles indicate votes in which the voting Democrats' proportion of ``Ayes'' is greater than the voting Republicans' proportion of ``Ayes'' whereas red triangles indicate votes in which the reverse happened.}
\label{fig:votespecificWAIC}
\end{figure}

We further explore the differences between our profit unfolding model and BGGUM during the 107\textsuperscript{th} House by computing vote-specific WAIC scores.  The associated score for vote $j$ undel model $m$ is defined analogously to \eqref{eq:lWAIC}:
\begin{multline*}
WAIC_j(m) = -2\left[\log\left( \textrm{E}_{\textrm{post}} \left\{ 
    \prod_{i = 1}^I \theta_{i,j}(m)^{y_{i,j}} \left[1 -\theta_{i,j}(m)\right]^{1 - y_{i,j}} \right\} \right) \right. \\
    \left. -  \textrm{var}_{\textrm{post}}\left\{ 
    \sum_{i = 1}^{I} \left[ y_{i,j} \log \theta_{i,j}(m) + (1-y_{i,j}) \log(1-\theta_{i,j}(m)) \right] \right\} \right].
\end{multline*}
These scores allow us to identify which votes are better explained by each of the two unfolding model (in terms of complexity-adjusted fit).  

Figure \ref{fig:votespecificWAIC} shows the difference between vote-specific WAIC scores under BGGUM and the probit unfolding model.  About 64\% of the votes during the 107\textsuperscript{th} House are better explained by our probit unfolding model. We also see a few votes where the difference of WAIC scores is large in either direction.  In particular, there are two votes in which BGGUM substantially outperforms the probit unfolding model.  Both of these correspond to Approval of the Journal (ApJ) votes.  Article I, section 5 of the U.S.\ Constitution requires that the House keep a journal of its proceedings, which is a summary of the day’s actions. The Speaker is responsible for examining and approving the Journal of the previous day. Following the announcement of approval by the Speaker, any House member may demand a vote on the question of the Speaker’s approval.  Journal-approval record votes rarely fail (of 472 held between 1991 and 2016, none failed), but various party and political considerations play a role in the decision to call for such a vote \citep{Hudiburg2018}.  Our reading of the literature suggests that voting patterns on Approval of the Journal votes are very idiosyncratic. In particular, \citet{patty2010dilatory} notes that votes on the Journal’s approval are just as frequently requested by the majority party as by members of the minority party, and that votes recorded on days on which a vote was also recorded on the House Journal are more likely to be close and more likely to be party-line votes than those recorded on other days.  Hence, we do not see the fact that BGGUM seems to explain these votes better as strong evidence in its favor.  On the other hand, the five highlighted votes that are better explained by the probit unfolding model are all substantive votes with response functions that are estimated to be non-monotonic by both models but where the ranks associated with the probit unfolding model are more consistent with an ends-against-the-middle voting pattern.


\begin{figure}[!t]
\centering
    \centering
    \includegraphics[width = 0.7\textwidth]{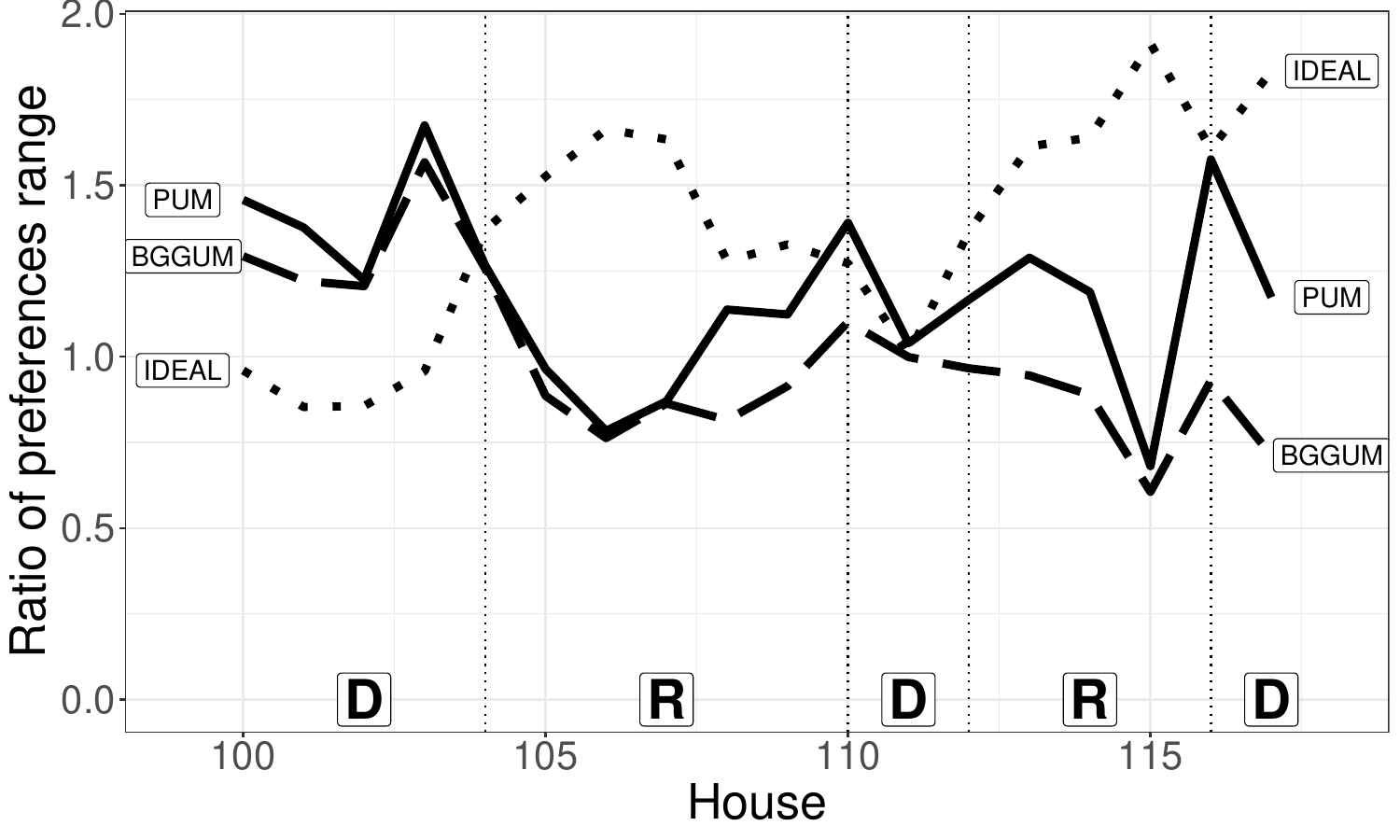}
\caption{Posterior mean of the ratio of Democrats' range over Republicans' range across the various Houses. The solid line corresponds to the probit unfolding model, the dotted line to IDEAL, and the dashed line to BGGUM.}
\label{fig:static_range_comp}
\end{figure}

The differences in the rankings of legislators we just discussed seem to be driven by substantial differences in the underlying estimates of the ideal points.  To illustrate this, we present in Figure \ref{fig:static_range_comp} the posterior mean of the ratio of the range of the ideal points of Republican and Democratic legislators for each of the three models under consideration.  Note that these ratios are invariant to affine transformations of the policy space, and are therefore identifiable and can be compared over time and across models. Generally speaking, we again see that the estimates for the probit unfolding model and for BGGUM are very similar to each other during the first half of the period under study, and quite different from those obtained through IDEAL.  The difference is particularly striking in the 100\textsuperscript{th}, 101\textsuperscript{st},  102\textsuperscript{nd} and 103\textsuperscript{rd} Houses, where the  unfolding models suggest that the ideological spread of Democrats is much larger than that of Republicans, and in the 105\textsuperscript{th}, 106\textsuperscript{th} and 107\textsuperscript{th}, where the estimates suggest the opposite.  On the other hand, starting with the 111\textsuperscript{th} House, the similarities between BGGUM and our probit unfolding model are less pronounced.  The behavior during the 115\textsuperscript{th} and 116\textsuperscript{th} Houses is particularly interesting.  In the 115\textsuperscript{th} House, our model agrees with BGGUM in estimating similar spreads for both parties.  This is in contrast to IDEAL, which estimates the spread of Democrats as almost twice as big as that of Republicans.  However, in the 116\textsuperscript{th} House (when Democrats regained control of the House), our model closely agrees with IDEAL in estimating a broader spread for Democrats. Section 3 of the Supplementary Materials presents alternative versions of Figure \ref{fig:static_range_comp} based on the standard deviation and the interquartile range of the estimated ideal points. The general patterns are very similar no matter what measure of dispersion is used, but some of the details do differ.  For example, the differences between PUM and the probit unfolding model that we observed for the range between the 112\textsuperscript{th} and 115\textsuperscript{th} House are less pronounced when using the standard deviation, and even less so for the interquartile range.
\begin{figure}[!t]
\centering
\begin{subfigure}[t]{.45\textwidth}
    \centering
    \includegraphics[width = .95\textwidth]{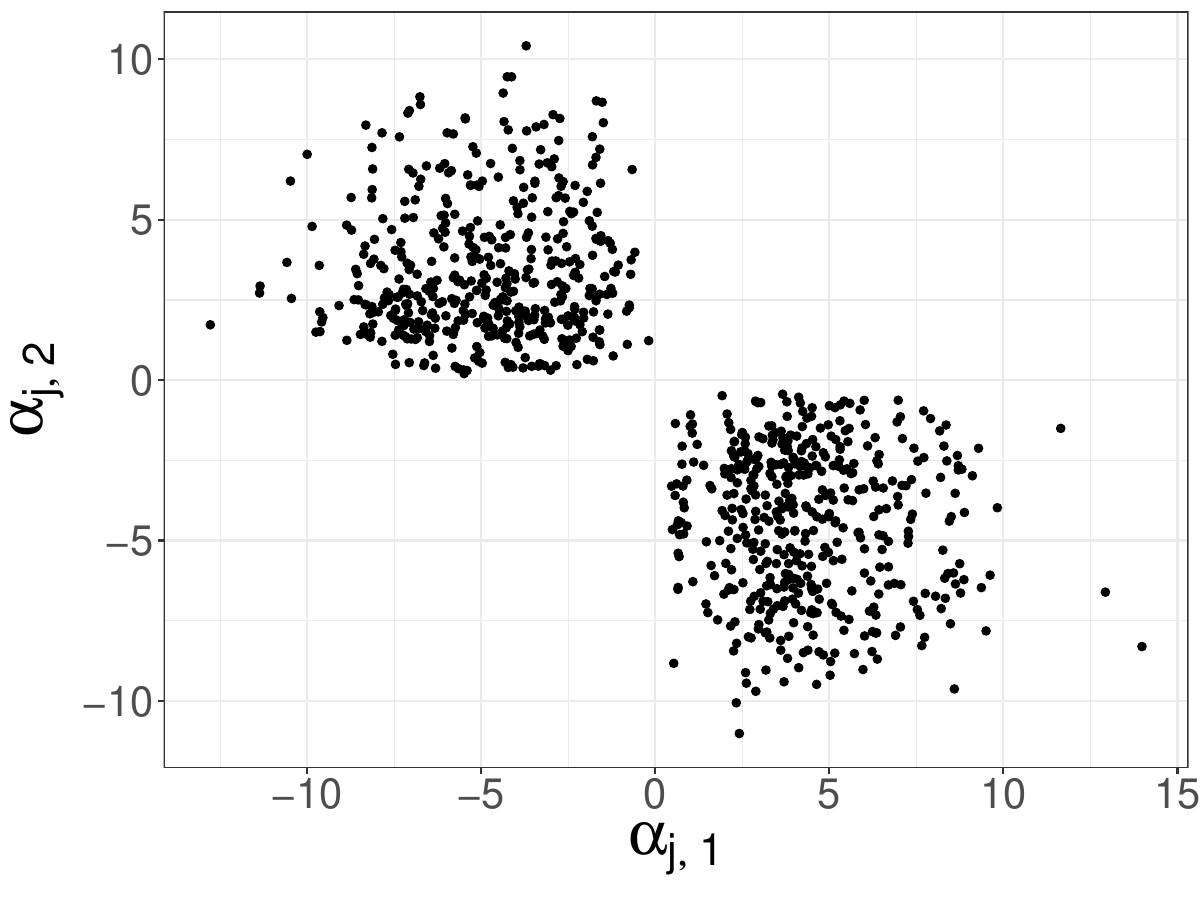}
    \caption{Posterior means}
\end{subfigure}
\begin{subfigure}[t]{.45\textwidth}
    \centering
    \includegraphics[width = .95\textwidth]{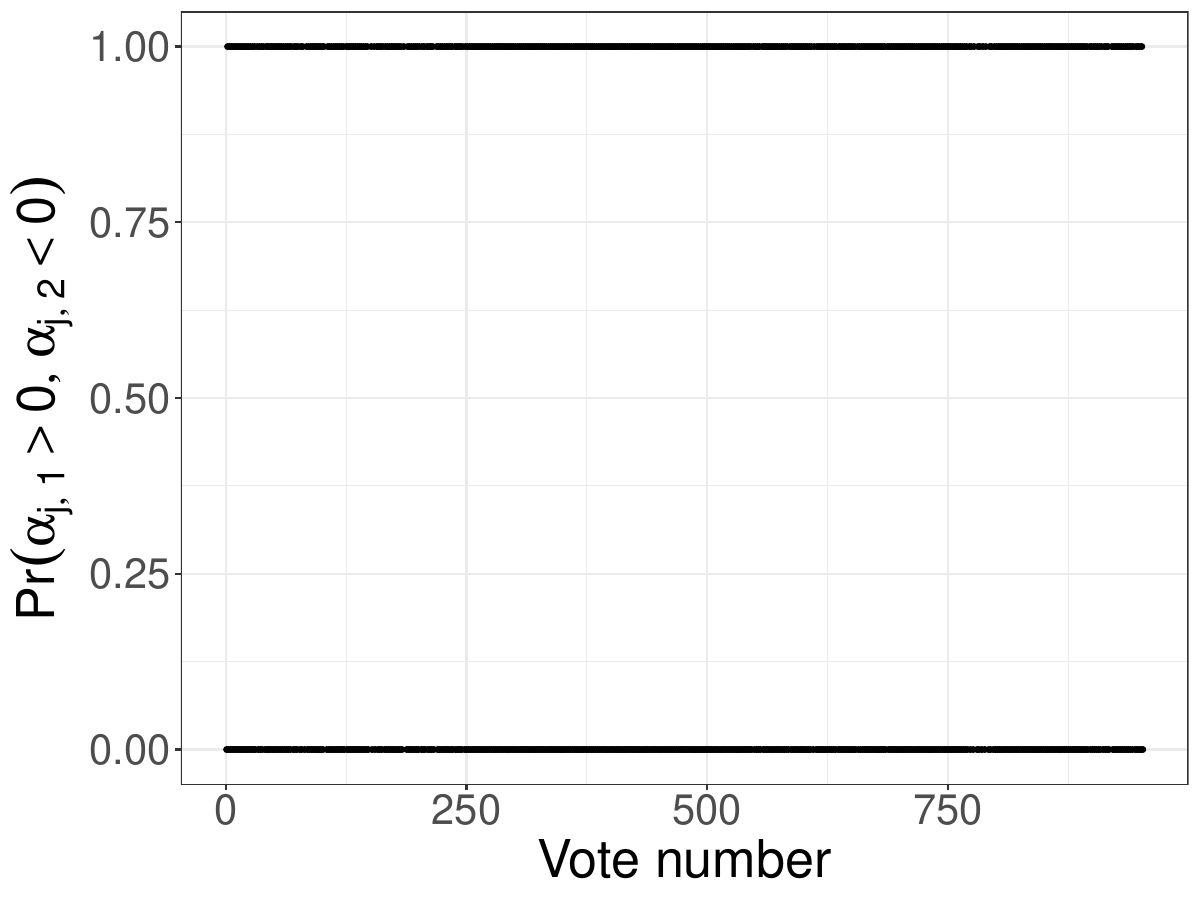}
    \caption{Posterior probability $\alpha_{j,1} > 0$ and $\alpha_{j,2} < 0$}
\end{subfigure}
\caption{Posterior summaries of $\bfalpha_{j}$ in the probit unfolding model for the 116\textsuperscript{th} House.}
\label{fig:alpha_plots_116}
\end{figure}
\begin{figure}[!b]
\centering
\begin{subfigure}[t]{.3\textwidth}
    \centering
    \includegraphics[width = .95\textwidth]{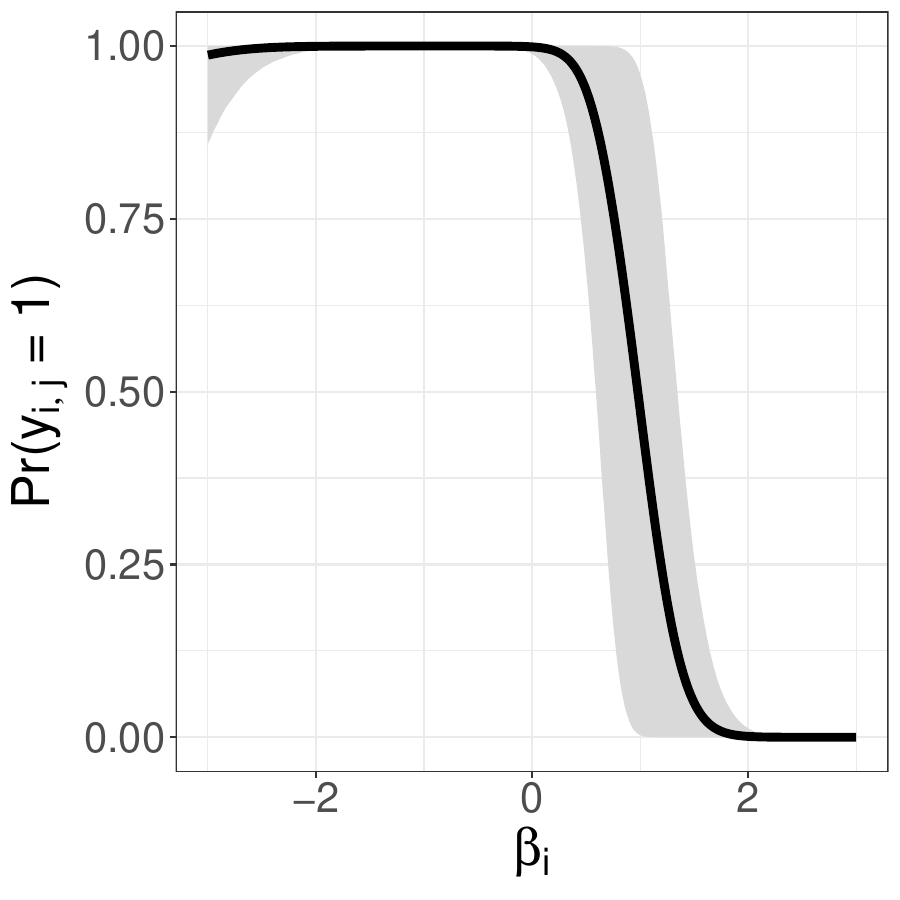}
    \caption{HRES5}
\end{subfigure}
\begin{subfigure}[t]{.3\textwidth}
    \centering
    \includegraphics[width = .95\textwidth]{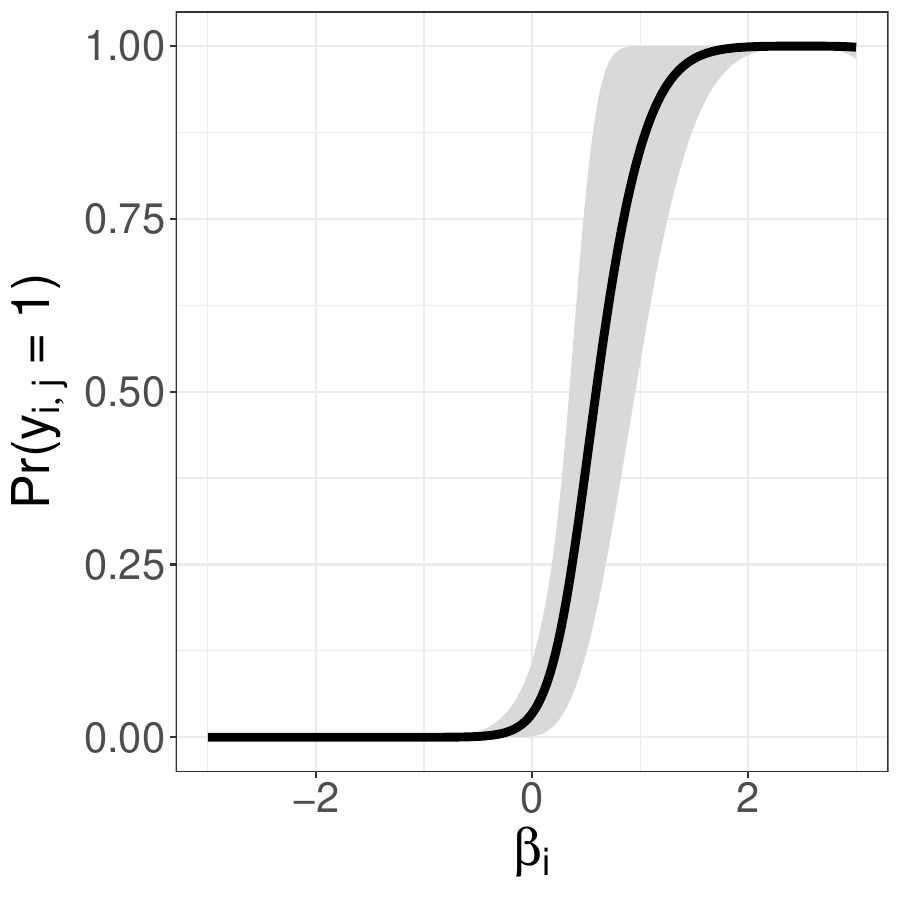}
    \caption{HR21}
\end{subfigure}
\begin{subfigure}[t]{.3\textwidth}
    \centering
    \includegraphics[width = .95\textwidth]{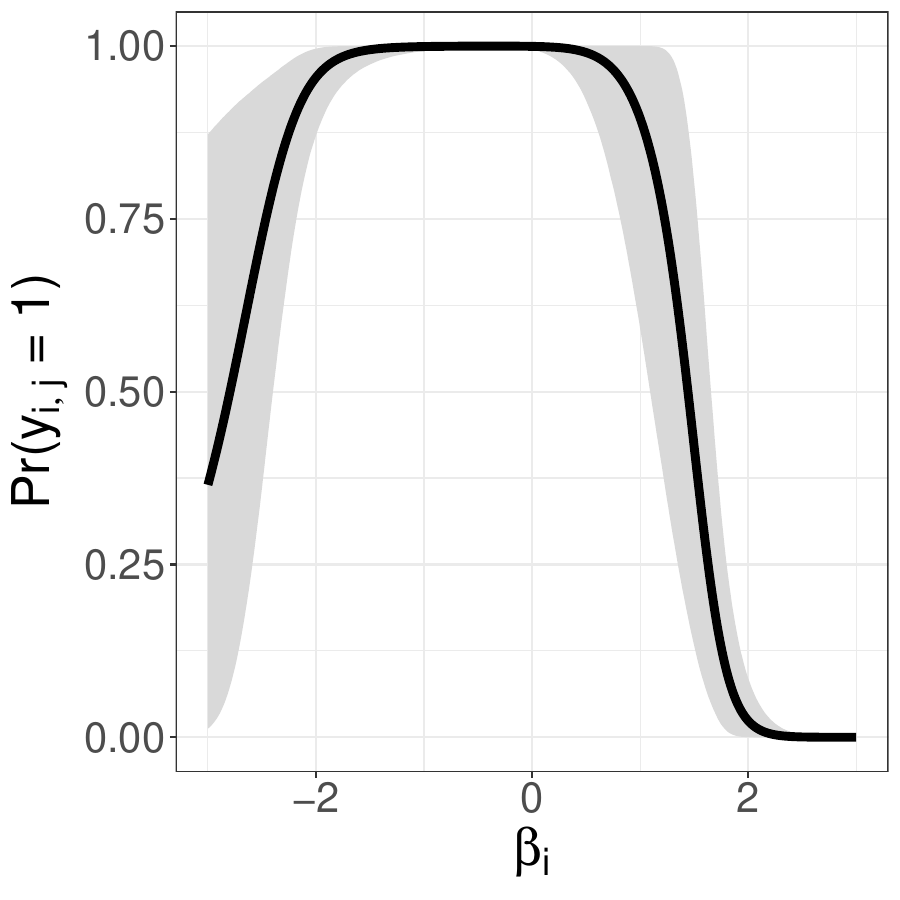}
    \caption{HRES6}
\end{subfigure}
\caption{Plots displaying various response curves based on the posterior means of $\alpha_{j,1}, \alpha_{j,2}$, $\delta_{j,1}$, and $\delta_{j,2}$ from the probit unfolding model for the 116th House. Here, the vote number refers to the clerk's roll call vote number.  Shaded areas correspond to 95\% pointwise posterior credible intervals.}
\label{fig:response_curves_116}
\end{figure}

Moving on now to a detailed analysis of the 116\textsuperscript{th} House, Figure \ref{fig:alpha_plots_116} shows posterior summaries for the distribution of $\bfalpha_1, \ldots, \bfalpha_J$.  The posterior distribution for 37\% of the $\bfalpha_j$'s in this House clearly lie on the lower right quadrant (where $\alpha_{j,1}>0$ and $\alpha_{j,2}<0$), while the remaining 63\% clearly lies on the upper left quadrant.  The structure of this posterior distribution provides support for our choice of a prior that allows for $\bfalpha$ to lie on either of the two quadrants, as opposed to a prior that concentrates on just one. 
On the other hand, Figure \ref{fig:response_curves_116} shows estimates of the response functions for three different votes.
These estimates demonstrate that the probit unfolding model is able to recover both monotonic and non-monotonic response functions from observed data. The left panel of Figure \ref{fig:response_curves_116} shows the response function for the first in a series of votes on HRES5, which provided for consideration of HRES6 (rules of the House of Representatives for the 116\textsuperscript{th} Congress) and HR21 (appropriations for the FY 2019). This response function is monotonically decreasing, which is to be expected given that the 116\textsuperscript{th} House had a Democratic majority. Indeed, recall from Section \ref{sec:identifiability} that the direction of latent space was identified by making the sign of the ideal point of the Republican whip positive.  This means that bills with a monotonically decreasing response function are bills favored by Democrats and not by Republicans.  Accordingly, all 230 Democrats present voted in favor of HRES5, while all 197 Republicans present voted against it.  Similarly, the center panel of Figure \ref{fig:response_curves_116} shows the response function for the first vote on HR21, a procedural vote proposed by Republicans.  The response function in this case is increasing, which agrees with the fact that all Democrats present voted against it while all Republicans voted in favor.  Finally, the right panel of Figure \ref{fig:response_curves_116} shows an example of a non-monotonic response function, corresponding to the first vote taken on HRES6 (rules of the House of Representatives for the 116\textsuperscript{th} Congress).  In this case the voting record is mixed:  most Democrats and three Republicans voted in favor, while most Republicans and three Democrats voted against the rules.  The disagreements within each party are, however, qualitatively different. The three Republican dissenters (Representatives Brian Fitzpatrick, John Katko and Tom Reed) are considered by most observers as moderates and have often expressed interest in bipartisanship.  On the other hand, the Democratic dissenters (Representatives Alexandria Ocasio-Cortez and Ro Khanna) are widely considered among the most liberal legislators in the House.  The shape of the response function reflects this:  the probability of a positive vote on this bill is high for most Democrats and for the more centrists Republicans, and lower for the more extreme legislators of either party.

\section{Dynamic unfolding models}\label{sec:dynamicmodel}

The analysis in the previous section relies on fitting the model described in Section \ref{sec:staticmodel} independently for each House.  While straightforward, this approach does not allow us to fully assess how individual members' preferences evolve over time. In this section, we consider an extension of the probit unfolding model that allows for a longitudinal analyses of member's preferences and extends the approach introduced in \citet{MartinQuinnDynamicIdealPoint2002a} to situations in which allowing for non-monotonic response functions might be desirable.

In a manner similar to Section \ref{sec:staticmodel}, let $y_{i, j, t}$ be the vote of member $i=1,\ldots,I$ on issue $j=1,\ldots, J_t$ considered on period $t=1, \ldots, T$.  As before, we postulate that voters make decisions based on three random utility functions,
\begin{align*}
    U_{N^{-}}(\beta_{i,t}, \psi_{j,t,1}) &= -\left(\beta_{i,t} - \psi_{j,t,1}\right)^2 + \epsilon_{i,j,t,1},\\
    U_{Y}(\beta_{i,t}, \psi_{j,t,2}) &= -\left(\beta_{i,t} - \psi_{j,t,2}\right)^2 + \epsilon_{i,j,t,2},\\
    U_{N^{+}}(\beta_{i,t}, \psi_{j,t,1}) &= -\left(\beta_{i,t} - \psi_{j,t,3}\right)^2 + \epsilon_{i,j,t,3},
\end{align*}
where $\epsilon_{i,j,t,1}$, $\epsilon_{i,j,t,2}$, and $\epsilon_{i,j,3}$ are independent and identically distributed Gaussian shocks, so that 
\begin{multline*}
    \textrm{P}(y_{i,j,t} = 1 \mid \beta_{i,t}, \alpha_{j,t,1}, \delta_{j,t,1}, \alpha_{j,t,2}, \delta_{j,t,2}) = \\ \int_{-\infty}^{\alpha_{j,t,1}(\beta_{i, t} - \delta_{j,t,1})} \int_{-\infty}^{\alpha_{j,t,2}(\beta_{i,t} - \delta_{j,t,2})} \frac{1}{2 \sqrt{3} \pi} \exp\left\{ -\frac{1}{3} (z_1^2 - z_1 z_2 + z_2^2) \right\} \dd z_1 \dd z_2.
\end{multline*}

We also continue treating the issue-specific parameters $(\bfalpha_{j,t}, \bfdelta_{j,t})$ as independent and identically distributed for all $j$ and $t$, and assign them the prior in Equation \eqref{eq:alphadeltaprior}.  The key difference with Section \ref{sec:staticmodel} lies on how we model the ideal points $\beta_{i,t}$.  Rather than completely independent priors for all $i$ and $t$, we treat the vectors $\bfbeta_1, \ldots, \bfbeta_I$ as independent, but assign each vector $\bfbeta_i = (\beta_{i,1}, \ldots, \beta_{i,T})'$ a joint Gaussian prior with mean $\mathbf{0}$ and covariance matrix
\begin{equation*}
\bfOmega(\rho) = \begin{pmatrix}
    1 & \rho & \rho^2 \ldots & \rho^{T-1} \\
    \rho & 1 & \rho & \cdots & \rho^{T-2} \\
    \rho^2 & \rho & 1 & \cdots & \rho^{T-3} \\
    \vdots & \vdots & \vdots & \ddots & \vdots \\
    \rho^{T-1} & \rho^{T-2} & \rho^{T-3} & \cdots & 1
\end{pmatrix}
\end{equation*}
where the parameter $0 \le \rho \le 1$ is unknown and needs to be learned from the data.

There are several ways in which this joint prior can be motivated.  For example, this prior can be obtained as the finite dimensional marginal of a zero-mean Gaussian process with an exponential covariance function. Alternatively, it can be motivated as a realization of a first order stationary autoregressive process with autocorrelation $\rho$ and with a standard normal distribution as its stationary distribution.  Both of these motivations make it clear that for any $t$, the marginal distribution of $\beta_{i,t}$ is the same distribution used in Section \ref{sec:staticmodel}.  Furthermore, the parameter $\rho$ controls how much information is borrowed over time; $\rho=1$ corresponds to a model in which preferences are assumed to be constant over time whereas $\rho=0$ simply leads back to fitting independent probit unfolding models for each $t$.  Consistent with \citet{MartinQuinnDynamicIdealPoint2002a}, in the context of the application described in the next section, we assume that preferences evolve relatively slowly and assign $\rho$ a normal prior with mean $\eta = 0.9$ and standard deviation $\lambda = 0.04$, which is truncated to the interval $[0,1]$.  We also conduct a sensitivity analysis to ascertain the impact of this choice on the inferences we draw from the model.  Please see Section 5 of the Supplementary Materials for further details.

\subsection{Computation}\label{sec:computation_dynamic}

The computational approach described in Section \ref{sec:computation} can be adapted to the dynamic unfolding model.  Most of the steps in the Markov chain Monte Carlo algorithm remain the same, with the main differences being in the sampling of the vector $\bfbeta_i$ and the need for an additional step related to sampling the correlation $\rho$.

Conditionally on latent vectors $\bfy^*_{i,j,t} = (y^*_{i,j,t,1}, y^*_{i,j,t,2}, y^*_{i,j,t,3})'$ defined similarly to Equation \eqref{eq:auxrepresentation}, the posterior distribution for the vector $\bfbeta_i$ reduces to a multivariate normal distribution with a sparse precision matrix.  Because of this, sampling from this joint distribution can be done quite efficiently, even for large $T$, by either exploiting algorithms for sparse linear algebra, or by building a forward-backward algorithm similar to the one employed in \citet{MartinQuinnDynamicIdealPoint2002a}.\footnote{The alternative, sampling from the full conditional of each $\beta_{i,t}$, leads to an algorithm that mixes extremely poorly and takes a very long time to explore the posterior distribution.  This is one of the challenges with trying to extend the approach of \citet{duck2022ends} to dynamic settings.}  As for sampling the autocorrelation $\rho$, we rely on a Metropolis-Hasting algorithm that uses a Gaussian random walk on $\Upsilon = \log (\rho/(1-\rho))$ as the proposal distribution.  The variance of the random walk is tuned to target a 40\% acceptance rate.  This is the only step in the algorithm that relies on this type of approach, making implementation relatively straightforward.  Details of this algorithm can be found in  Section 2 of the Supplementary Materials, and code implementing the algorithm is available at \url{https://github.com/rayleigh/probit_unfolding_model/}.

\section{Revealed preferences in the U.S.\ Supreme Court, 1937-2021}\label{sec:resultsSCOTUS}

We use the dynamic model from Section \ref{sec:dynamicmodel} to examine the voting record of U.S. Supreme Court (SCOTUS) justices between 1937 to 2021.  We compare our results to those generated by the dynamic factor model described in \citet{MartinQuinnDynamicIdealPoint2002a}, which extends IDEAL to dynamic settings and has become the de-facto gold standard method for measuring justice's preferences.

The data we analyze is available at \url{https://mqscores.lsa.umich.edu/}.  As in  \citet{MartinQuinnDynamicIdealPoint2002a}, a justice's vote is encoded as 1 if they voted to reverse a lower court's decision, and 0 if they voted to affirm that decision.  Our inferences are based on 20,000 samples obtained after burning the first 300,000 iterations of our Markov chain Monte Carlo algorithm and thinning the next 200,000 every tenth observation. 
Results for the model from \citet{MartinQuinnDynamicIdealPoint2002a} (MQ in the sequel) were obtained using the \texttt{R} package \texttt{MCMCpack}.  We use 10,000 samples from the associated Markov chain Monte Carlo algorithm, obtained after burning the first 40,000 draws, from four chains. This gives us a total of 40,000 samples.
\begin{figure}[!t]
    \centering
    \begin{subfigure}[t]{0.49\textwidth}
        \centering
        \includegraphics[width = \textwidth]{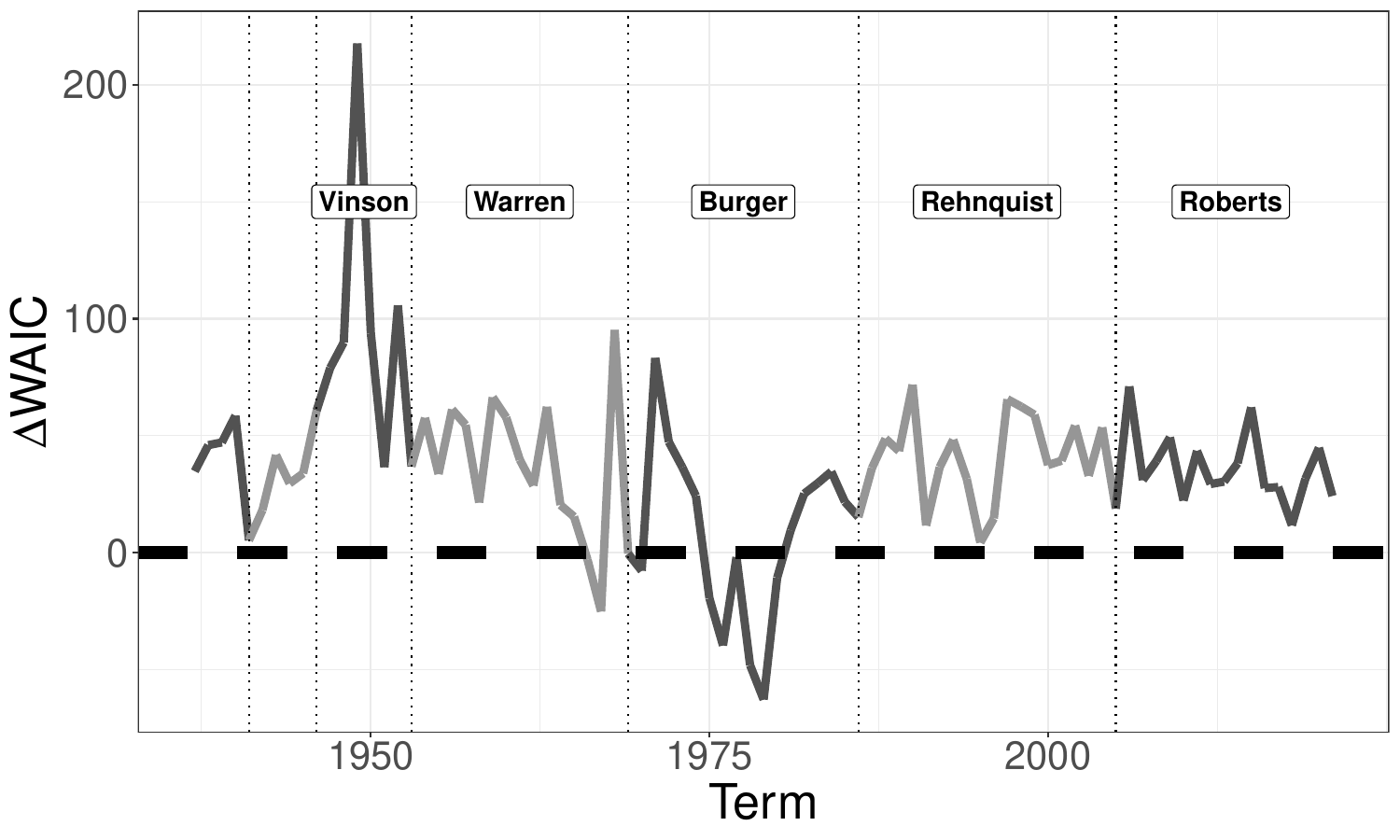}
        \caption{WAIC}\label{fig:WAIC_comparison_SCOTUS}
    \end{subfigure}
    \begin{subfigure}[t]{0.49\textwidth}
        \centering
        \includegraphics[width = \textwidth]{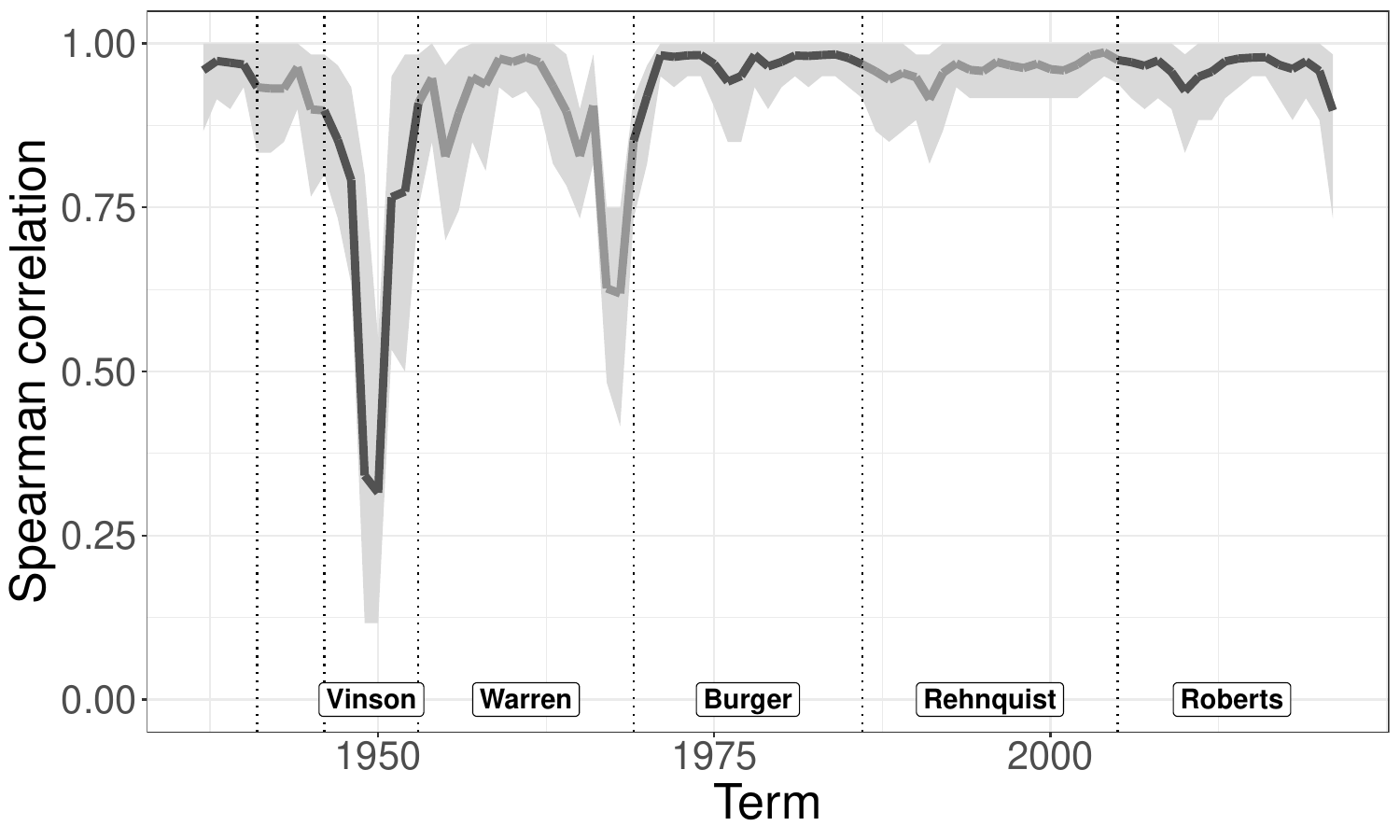}
        \caption{Spearman correlation}\label{fig:rank_comparison_SCOTUS}
    \end{subfigure}
\caption{Left panel: Difference in WAIC scores between MQ and the dynamic unfolding model ($WAIC(\text{MQ}) - WAIC(\text{DPUM})$).  Note that the way the difference is being computed here is the opposite to the way in which it was computed in Figure \ref{fig:static_waic_comp}.  Right panel:  Posterior mean (solid line) and corresponding 95\% credible intervals (shaded region) for the Spearman correlation between the justices' rankings generated by the dynamic unfolding model and MQ.}
\end{figure}

Figure \ref{fig:WAIC_comparison_SCOTUS} presents the difference in WAIC scores between MQ and the dynamic unfolding model for each of the terms over consideration (recall Equation \eqref{eq:lWAIC}, and note that the difference is being computed in the opposite way to Figure \ref{fig:static_waic_comp}). Under this metric, the dynamic unfolding model seems to consistently outperform MQ.  The exceptions are the 1966 -- 1967, 1969 -- 1970, and 1975--1980 terms, where MQ seems to dominate. Complementing these results, Figure \ref{fig:rank_comparison_SCOTUS}, presents the posterior mean and corresponding 95\% credible intervals for the Spearman correlation between the justices' rankings generated by the dynamic probit unfolding model and MQ. We can see that, in spite of the probit unfolding model dominating MQ in terms of WAIC scores, their rankings of the justices agree considerably, especially over the last 40 years.  Nevertheless, there are two periods during which the models drastically disagree:  the 1949-1952 and 1967 to 1970 periods.  

\begin{figure}[!b]
\centering
\begin{subfigure}[t]{0.45\textwidth}
    \centering
    \includegraphics[width = .95\textwidth]{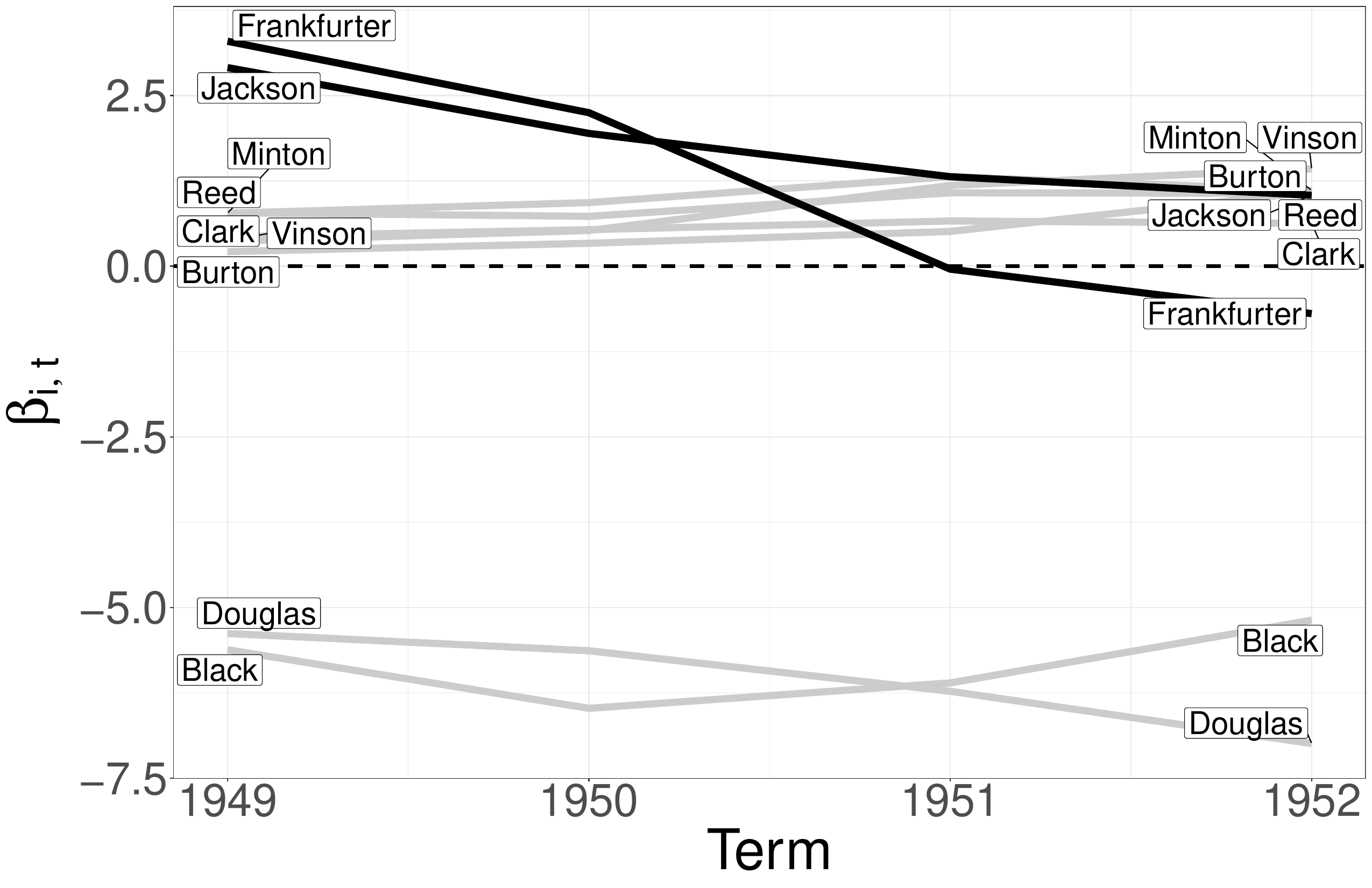}
    \caption{Dynamic probit unfolding model}\label{fig:scores1949to1952threeup}
\end{subfigure}
\begin{subfigure}[t]{0.45\textwidth}
    \centering
    \includegraphics[width = .95\textwidth]{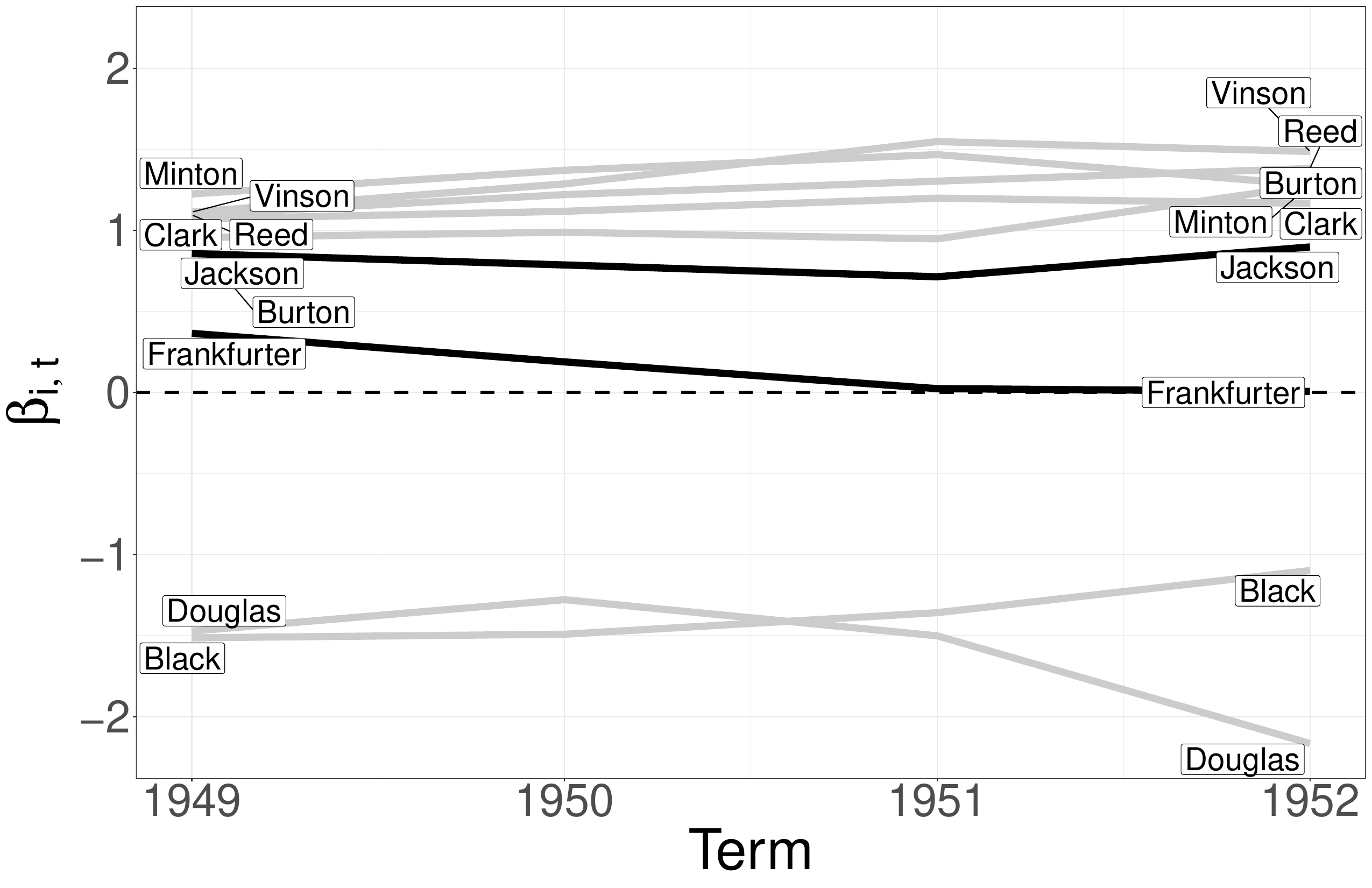}
    \caption{MQ}\label{fig:scores1949to1952MQ}
\end{subfigure} 
\caption{Posterior mean of the ideal points for SCOTUS Justices active during 1949 to 1952 terms under the dynamic unfolding model (left column) and MQ (right column).
} \label{fig:scores1949to1952}
\end{figure}

To be concise, we focus the rest of our discussion on the 1949-1952 period, which is also a period when the WAIC score greatly favors our dynamic unfolding model.  Between 1949 and 1952, SCOTUS was composed of five appointees of president Franklin D.\ Roosevelt (Justices Felix Frankfurter, Robert Jackson, Hugo Black, Willian O.\ Douglas and Stanley F.\ Reed), joined by four more recent appointees of Harry S.\  Truman (Justices Harold H.\ Burton, Fred M.\ Vinson, Tom C.\ Clark and Sherman Minton, with Clark and Minton serving their first term in 1949).  Figure \ref{fig:scores1949to1952} shows the posterior mean of the ideal points for these Justices under the dynamic unfolding model (left column) and MQ (right column).  We can see that the different rankings are driven by substantial differences in the estimates of Justices Frankfurter and Jackson preferences:  MQ places these two justices as centrists, while our dynamic unfolding model places them as the two most conservative members of the court in 1949 and 1950.  We argue that the characterization coming out of the dynamic unfolding model is in closer agreement with the accepted understanding of these Justices' ideological leanings.  Indeed, Justice Frankfurter represents a bit of an enigma for historians and legal scholar, but most have come to see him as the leader of the conservative faction of the Supreme Court (e.g., see \citealp{eisler1993justice}), to which Jackson (a frequent ally of Frankfurter) also belonged.  During his long career, first as a professor at Harvard and then as a Justice, Frankfurter was a staunch supporter of judicial restraint.  During the 1920s and early 1930s, liberals embraced judicial restraint as a way to check the power of the conservative justices that dominated the Supreme Court, placing Frankfurter firmly on the left wing of the judicial community.  However, as Roosevelt set to remake the Supreme Court, most liberals (but not Frankfurter) abandoned judicial restraint and embraced judicial activism.  This, along with his difficult relationship with other members of the court, explains how one of the minds behind Roosvelt's New Deal legislation and the creation of the American Civil Liberties Union came to be considered a stalwart of the conservative wing of the Supreme Court.

\section{Discussion}\label{sec:discussion}

This paper introduced a new class of unfolding models for binary preference data that can be motivated from first principles as a spatial voting model.  We also consider extensions to dynamic models that allow the preferences of legislators to evolve over time.  A key feature of this class of models is that it allows for non-monotonic response functions, which can  arise in practice when legislators at the extremes of the political spectrum vote together against the center. Our extensive evaluations on voting data from the U.S.\ House of Representatives and the U.S.\ Supreme Court indicate that the model substantially outperforms both traditional scaling models and alternative unfolding models that had been previously introduced in the literature.  

Our model is slightly more flexible than BGGUM, as it does not constrain the  discrimination parameters $\alpha_{j,1}$ and $\alpha_{j,2}$ (recall Equation \eqref{eq:logisticBGUM} and the associated discussion).  Of course, having a model with more parameters is not always advantageous.  However, the values of the WAIC in Sections \ref{sec:resultsUSHouse} and \ref{sec:resultsSCOTUS} indicate that, in the vast majority of cases, the additional flexibility of the probit unfolding model improves the overall fit enough to compensate for the slight increase in model complexity.  Furthermore, our results suggest that, while both BGGUM and our model sometimes misclassify relative centrists as extremists (or vice versa!), BGGUM tends to do so much more often. In fact, the work presented in this manuscript bears on the question of how to estimate \textit{ideological} rankings from voting data.  While the term \textit{ideology} has a long and varied history of usage in scholarship (e.g., see \citealp{gerring1997ideology}), it is most often used to refer to specific policy views and preferences held by individuals, either ``an underlying philosophy on which all specific political views are based'' (\citealp[page 17]{jessee2012ideology}) or a belief system that includes a wide range of opinions consistently held \citep{converse1964nature}.  However, the term is commonly operationalized in terms of revealed preferences and ideal points, as estimated by spatial voting models (e.g., \citealp{poole2006ideology}).  How faithful this operationalization is to the more etymological definitions is open to debate.  One key observation from our studies is that our unfolding model not only leads to superior complexity-adjusted fit measures such as the WAIC, but it also yields vote-based estimates of preferences that seem to better match to what, on the basis of the public record, would seem to be the true philosophy / belief system of legislators and justices.

Moving forward, there are several extensions of the framework introduced in this paper that we would like to to pursue. One refers to the class of link functions used to define the model. Instead of using a probit link, we might instead be interested in a logit link, which is more robust to outlier votes. Computation in this case is potentially more challenging, but the algorithm described in this paper could be adapted by using a mixture approximation for the Gumbel distribution (e.g., see \citealp{Fruhwirth-SchnatterFruhwirthBayesianInferenceMultinomial2012}). A second extension refers to the use of higher-dimensional policy spaces.  Right now, all unfolding models that we are aware of assume that the underlying latent space is unidimensional.  Our framework provides a natural setting in which to build higher dimensional models in a principled fashion.  Finally, we are interested in extending our framework to general multinomial ordinal observations. Such extension can be achieved by introducing pairs of issue-specific coordinates for each level of the observed categorical variables.

\begin{acknowledgement}
We would like to thank Kevin Quinn for his help with the \texttt{R} package \texttt{mcmcPack}.
\end{acknowledgement}

\paragraph{Funding Statement}

This research was supported by grants from the U.S.\ National Science Foundation DMS/CISE-2023495 and DMS-2114727.

\paragraph{Competing Interests}

None.


\nocite{MartinAndrewD.andKevinM.Quinn.2021MQScores2022}
\nocite{YuXingchenandAbelRodriguezSpatialVotingModels}
\nocite{LewisJeffreyB.KeithPooleHowardRosenthalAdamBocheAaronRudkinandLukeSonnetVoteviewCongressionalRollCall2023}
\printbibliography
\end{document}


\section{Details of the Markov chain Monte Carlo algorithm for the static probit unfolding model}\label{ap:MCMCdetails}

As discussed in Section 2.3 of the original manuscript, we consider a data augmentation approach where
\begin{align}
    \begin{pmatrix}
        y^*_{i, j, 1}\\
        y^*_{i, j, 2}\\
        y^*_{i, j, 3}\\
    \end{pmatrix}
    \bigg|
    \alpha_{j,1}, \alpha_{j,2}, \delta_{j,1}, \delta_{j,2}, \beta_i
    &\sim
    \textrm{N}\left(\begin{pmatrix}
        -\alpha_{j,1}(\beta_i - \delta_{j,1})\\
        0\\
        -\alpha_{j,2}(\beta_i - \delta_{j,2})\\
    \end{pmatrix} 
    \bigg| 
    \v{0}_3, 
    \begin{pmatrix}
        1 & 0 & 0\\
        0 & 1 & 0\\
        0 & 0 & 1\\
    \end{pmatrix}\right) ,
    \label{eq:utility_distr2}
\end{align}
for all $i=1,\ldots, I$ and $j=1,\ldots, J$, and let $y_{i, j} = 1$ if and only if $y^*_{i, j, 2} > \max \left\{ y^*_{i, j, 1}, y^*_{i, j, 3} \right\}$ and $y_{i, j} = 0$ otherwise. For notational convenience, we will let $\v{\widetilde{y}^*_{i, j}} = \{y^*_{i, j, 1}, y^*_{i, j, 3}\}$.

Additionally, because of the mixture structure associated with the prior on $\bfalpha_j$, $\bfdelta_j$,  we consider a second data augmentation scheme and introduce variables $z_1, \ldots, z_J$ such that $z_{j} = 1$ if and only if $\alpha_{j,1} > 0$ and $\alpha_{j,2} < 0$, and $z_{j} = -1$ otherwise, where $\Pr(z_i=1) = 0.5$ a priori. 

\paragraph{\textbf{Sampling $\beta_i$, $i = 1, 2, \ldots, I$ :}}  Conditional on $\v{y^*_{i, j}}$'s, $\bfalpha_{j}$'s and $\bfdelta_{j}$'s, the posterior distribution of each $\beta_i$ is Gaussian with mean $\mu_{\beta_i}$, and variance $\sigma^2_{\beta_i}$ given by:
\begin{align}
    \sigma^2_{\beta_i} = \frac{1}{1 + \sum_{j = 1}^{J} \v{\alpha_j}' \v{\alpha_j}} & &
    \mu_{\beta_i} = -\sigma^2_{\beta_i}\left(\sum_{j = 1}^J \v{\alpha_j}' (\v{\widetilde{y}^*_{i, j}} - D_{\v{\alpha_j}}\v{\delta_i})\right).
\end{align}

\paragraph{\textbf{Sampling $z_j$, $j = 1, 2, \ldots, J$ :}}  Given $\v{y^*_{i, j}}$s, $\beta_i$s $\bfdelta_{j}$, and after integrating over $\bfalpha_j$, the conditional for $z_j=1$ is proportional to the following:
\begin{align}
    p(z_j = 1 \mid \v{y^*_{i, j}} \v{\delta_j}, \{\beta_i\}) \nonumber \;\;\;\;\;\;\; \;\;\;\;\;\;\; \;\;\;\;\;\;\; \;\;\;\;\;\;\; \;\;\;\;\;\;\; \;\;\;\;\;\;\; \;\;\;\;\;\;\; \;\;\;\;\;\;\; \;\;\;\;\;\;\; \;\;\;\;\;\;\; \;\;\;\;\;\;\; \;\;\;\;\;\;\; \;\;\;\;\;\;\; \;\;\;\;\;\;\; \;\;\; \\
    \propto \phi\left(\v{\delta_j} \mid \v{\mu}, \kappa^2 \mathbbm{I}_{2\times 2}\right) \int_{0}^{\infty} \int_{-\infty}^{0} \prod_{i} p(\v{y^*_{i, j}} \mid \v{\alpha_j}, z_j = 1, \v{\delta_j}, \beta_i) p(\v{\alpha_j} \mid z_j = 1) d \alpha_{j,2} d \alpha_{j,1}  \nonumber \\
    = \phi \left(\v{\delta_j} \mid \v{\mu}, \kappa^2 \mathbbm{I}_{2\times 2}\right)  \left\{ 1 - \Phi\left(  -(\mu_{\alpha_j})_1 / \left(\Sigma_{\alpha_j}\right)_{1, 1} \right) \right\} \Phi\left(- (\mu_{\alpha_j})_2 / \left(\Sigma_{\alpha_j}\right)_{2, 2} \right) ,
\end{align}
where $\phi(\cdot \mid \bfmu, \bfSigma)$ is the density of a (multivariate) Gaussian distribution with mean $\bfmu$ and variance $\bfSigma$, $\Phi$ the  cumulative distribution function of the (univariate) standard Gaussian distribution, $\v{\mu_{\alpha_j}}$ and $\Sigma_{\alpha_j}$ are defined as
\begin{align}
    \Sigma_{\alpha_j} = \left(\sum_{i} D_{\beta_i - \v{\delta_j}} D_{\beta_i - \v{\delta_j}} + \frac{1}{\omega^2}\mathbbm{I}_{2, 2}\right)^{-1} & &
    \mu_{\alpha_j} =  
    - \Sigma_{\alpha_j} \left(\sum_{i} D_{\beta_i - \delta}  \v{\widetilde{y}^{*}_{i, j}}\right),
    \label{eq:alpha_post_mu_cov}
\end{align}
with $D_{(a, b)}$ indicating a diagonal matrix with diagonal entries $a$ and $b$ for $a, b \in \mathbbm{R}$, and $(\mu_{\alpha_j})_k$ and $\left(\Sigma_{\alpha_j}\right)_{k, k'}$ represent the $k$-th and $(k,k')$-th entries of $\mu_{\alpha_j}$ and $\Sigma_{\alpha_j}$, respectively.

Similarly, the conditional for $z_j = -1$ is given by 
\begin{align}
    p(z_j = -1 \mid \v{y^*_{i, j}} \v{\delta_j}, \{\beta_i\}) \nonumber \;\;\;\;\;\;\; \;\;\;\;\;\;\; \;\;\;\;\;\;\; \;\;\;\;\;\;\; \;\;\;\;\;\;\; \;\;\;\;\;\;\; \;\;\;\;\;\;\; \;\;\;\;\;\;\; \;\;\;\;\;\;\; \;\;\;\;\;\;\; \;\;\;\;\;\;\; \;\;\;\;\;\;\; \;\;\;\;\;\;\; \;\;\;\;\;\;\; \;\;\; \\
    \propto \phi\left(\v{\delta_j} \mid -\v{\mu}, \kappa^2 \mathbbm{I}_{2\times 2}\right) \int_{-\infty}^{0} \int_{0}^{\infty} \prod_{i} p(\v{y^*_{i, j}} \mid \v{\alpha_j}, z_j = 1, \v{\delta_j}, \beta_i) p(\v{\alpha_j} \mid z_j = -1) d \alpha_{j,2} d \alpha_{j,1}  \nonumber \\
    = \phi \left(\v{\delta_j} \mid -\v{\mu}, \kappa^2 \mathbbm{I}_{2\times 2}\right)   \Phi\left(  -(\mu_{\alpha_j})_1 / \left(\Sigma_{\alpha_j}\right)_{1, 1} \right) \left\{ 1 - \Phi\left(- (\mu_{\alpha_j})_2 / \left(\Sigma_{\alpha_j}\right)_{2, 2} \right) \right\} .
\end{align}

\paragraph{\textbf{Sampling $\bfalpha_j$, $j = 1, 2, \ldots, J$ :}} The posterior distributions of $\bfalpha_{j}$ conditioned on $z_j$, $\beta_i$s, $\v{y^*_{i, j}}$s and $\bfdelta_{j}$ is a truncated multivariate normal distributions for $j = 1, 2, \ldots, J$. The posterior mean for $\alpha_{j,1}$ and $\alpha_{j,2}$, $\v{\mu_{\alpha_j}}$, and posterior covariance matrix, $\Sigma_{\alpha_j}$, are given in \eqref{eq:alpha_post_mu_cov}. Meanwhile, the truncation is determined by $z_j$.  If $z_j=1$, then the distribution is truncated to the region where $\alpha_{j,1}>0$ and $\alpha_{j,2}<0$, and it is truncated to the region $\alpha_{j,1}<0$ and $\alpha_{j,2}>0$ if $z_j=-1$. 

Note that we sample $z_j$ and $\v{\alpha_j}$ in the same Gibbs step because we compute the posterior of $\v{\alpha_j}$ when sampling for $z_j$. Of course, we perform this Gibbs step by sampling $z_j$ conditioned on $\beta_i$'s and $\v{\delta_j}$ and then sampling $\v{\alpha_j}$ conditioned on $\beta_i$'s, $z_j$, and $\v{\delta_j}$.

\paragraph{\textbf{Sampling $\bfdelta_j$, $j = 1, 2, \ldots, J$ :}} Finally, the posterior distributions of $\bfdelta_{j}$ conditioned on $z_j$, $\beta_i$'s, $\v{y^*_{i, j}}$'s, and $\bfalpha_{j}$ is a multivariate normal distributions with posterior mean  $\v{\mu_{\delta_j}}$ and posterior covariance matrix $\Sigma_{\delta_j}$ given by:
\begin{align}
    \Sigma_{\delta_j} = \left(I D_{\v{\alpha_j}} D_{\v{\alpha_j}} + \frac{1}{\kappa^2}\mathbbm{I}_{2, 2}\right)^{-1} & &
    \mu_{\delta_j} = \Sigma_{\delta_j} \left(\sum_{i = 1}^I D_{\v{\alpha_j}} (\v{\widetilde{y}^{*}_{i, j}} - \v{\alpha_j}\beta_i) + z_j \frac{\v{\mu}}{\kappa^2}\right).
\end{align}

\paragraph{\textbf{Sampling $\v{y^*_{i, j}}$; $i = 1, 2, \ldots, I$ $j = 1, 2, \ldots, J$ :}} Conditioned on $\v{\alpha_j}$, $\v{\delta_j}$, and $\beta_i$, we first discuss how to sample $\v{y^*_{i, j}}$ if $y_{i,j} = 1$.
\begin{enumerate}
    \item Draw $y^*_{i, j, 1}$ from a normal distribution with mean $-\alpha_{j,1}(\beta_i - \delta_{j,1})$ and variance 1 truncated to the interval, $(-\infty, y^*_{i, j, 2})$. 
    \item Draw $y^*_{i, j, 2}$ from a standard normal distribution truncated to the interval, $(\max(y^*_{i, j, 1}, y^*_{i, j, 3}), \infty)$.
    \item Draw $y^*_{i, j, 3}$ from a normal distribution with mean $-\alpha_{j,2}(\beta_i - \delta_{j,2})$ and variance 1 truncated to the interval, $(-\infty, y^*_{i, j, 2})$. 
\end{enumerate}

Next, we discuss how to sample $\v{y^*_{i, j}}$ if $y_{i,j} = 0$.
\begin{enumerate}
    \item If $y^*_{i, j, 3} > y^*_{i, j, 2}$, draw $y^*_{i, j, 1}$ from a normal distribution with mean $-\alpha_{j,1}(\beta_i - \delta_{j,1})$ and variance 1. Otherwise, draw $y^*_{i, j, 1}$ from the same normal distribution, but truncate this normal distribution to the interval $(y^*_{i, j, 2}, \infty)$. 
    \item Draw $y^*_{i, j, 2}$ from a standard normal distribution truncated to the interval, $(-\infty, \max(y^*_{i, j, 1}, y^*_{i, j, 3}))$.
    \item If $y^*_{i, j, 1} > y^*_{i, j, 2}$, draw $y^*_{i, j, 3}$ from a normal distribution with mean $-\alpha_{j,2}(\beta_i - \delta_{j,2})$ and variance 1. Otherwise, draw $y^*_{i, j, 3}$ from the same normal distribution, but truncate this normal distribution to the interval $(y^*_{i, j, 2}, \infty)$
\end{enumerate}



\section{Details of the Markov chain Monte Carlo algorithm for the dynamic probit unfolding model}\label{ap:MCMCdetails2}

Similarly to Section \ref{ap:MCMCdetails}, we augment the sampler with variables
\begin{align}
    \begin{pmatrix}
        y^*_{i, j, t, 1}\\
        y^*_{i, j, t, 2}\\
        y^*_{i, j, t, 3}\\
    \end{pmatrix}
    \bigg|
    \alpha_{j,t,1}, \alpha_{j,t,2}, \delta_{j,t,1}, \delta_{j,t,2}, \beta_{i,t}
    &\sim
    \textrm{N}\left(\begin{pmatrix}
        -\alpha_{j,t,1}(\beta_{i,t} - \delta_{j,t,1})\\
        0\\
        -\alpha_{j,t,2}(\beta_{i,t} - \delta_{j,t,2})\\
    \end{pmatrix} \bigg| 
    \begin{pmatrix}
        1 & 0 & 0\\
        0 & 1 & 0\\
        0 & 0 & 1\\
    \end{pmatrix}\right) ,
    \label{eq:utility_distr}
\end{align}
where $y_{i, j, t} = 1$ if and only if $y^*_{i, j, t, 2} > \max \left\{ y^*_{i, j, t, 1}, y^*_{i, j, t, 3} \right\}$ and $y_{i, j, t} = 0$ otherwise, and also define
$z_{j,t} = 1$ if $\alpha_{j,t,1} > 0$ and $\alpha_{j,t,2} < 0$, and $z_{j,t} = -1$ otherwise, where $\Pr(z_{i,t}=1) = 0.5$ a priori. 

The steps associated with sampling the $\bfdelta_{j,t}$s, $\bfalpha_{j,t}$s, $z_{j,t}$s and $\bfy_{i,t,j}^*$s are analogous to those described in the previous section.  We describe the steps for sampling $\bfbeta_{t}$ and $\rho$ below:

\begin{figure}[!t]
\centering
\begin{subfigure}[t]{.45\textwidth}
    \centering
    \includegraphics[width = .95\textwidth]{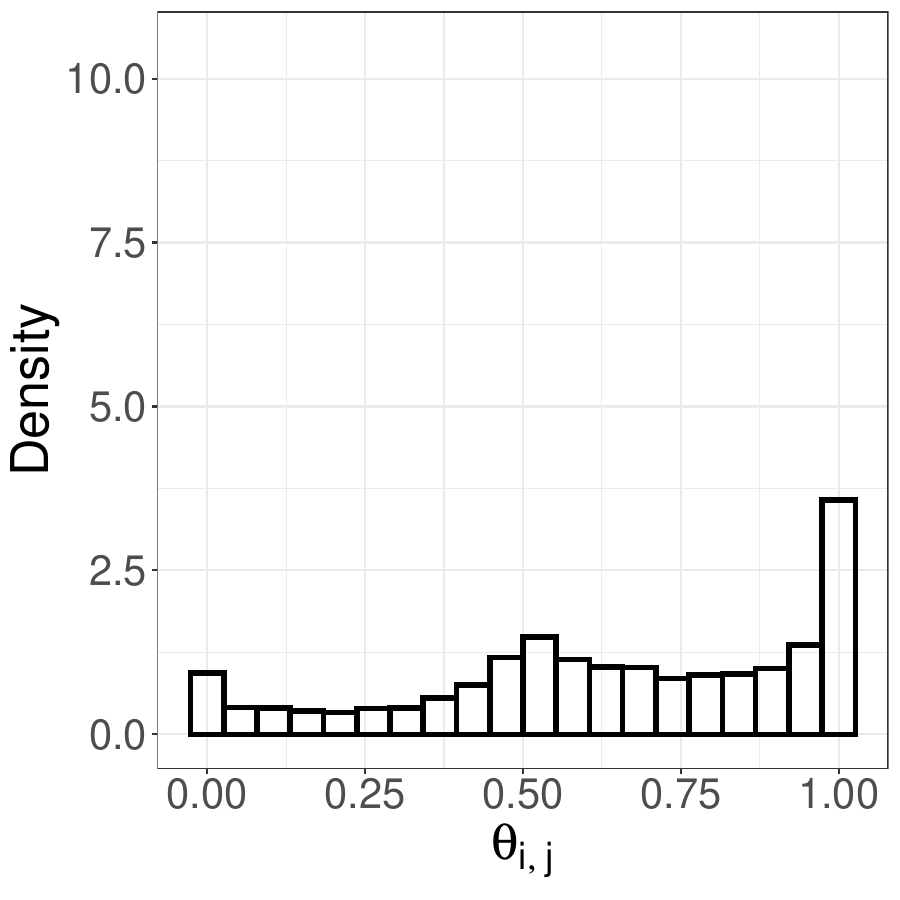}
    \caption{Alternative prior}
\end{subfigure}
\begin{subfigure}[t]{.45\textwidth}
    \centering
    \includegraphics[width = .95\textwidth]{img/overview/yea_prob_prior_plot.pdf}
    \caption{Original prior}
\end{subfigure}
\caption{Histograms for 10,000 draws of the implied prior distribution on $\theta_{i,j}$ for our probit GGUM model under an alternative prior with $\bfmu = (-2,10)'$, $\omega^2 = 1$ and $\kappa^2 = 9$ (left panel) and under the original prior with  $\bfmu = (-2,10)'$, $\omega^2 = 25$ and $\kappa^2 = 10$.}
\label{fig:prior_draws_prior_comp}
\end{figure}

\paragraph{\textbf{Sampling $\v{\beta_i}$, $i = 1, 2, \ldots, I$ :}}  To ensure proper mixing of our Markov chain, we sample the whole trajectory of ideal points for the $i$-th justice, $\v{\beta_i}$, from its joint full conditional distribution.  This corresponds to a multivariate normal distribution with variance matrix $\Sigma_{\beta_i} = \left\{ B + \bfOmega(\rho)^{-1} \right\}^{-1}$ and mean $\mu_{\beta_i} = -\Sigma_{\beta_i} \mathbf{m}$,
where $\v{m}$ is a vector with entries 
\begin{align}
   m_t &= -\sum_{j : y_{i, j, t} \in \{0, 1\}} \left(\alpha_{j, t, 1}(y^*_{i, j, 1} - \alpha_{j, t, 1} \delta_{j, 1}) + \alpha_{j, t, 2}(y^*_{i, j, 3} - \alpha_{j, t, 2} \delta_{j, t, 2})\right)
\end{align}
and $\mathbf{B}$ is a diagonal matrix with entries
\begin{align}
   B_{t, t} &= \sum_{j : y_{i, j, t} \in \{0, 1\}} \v{\alpha}'_{j, t} \v{\alpha}_{j, t}. 
\end{align}

\paragraph{\textbf{Sampling $\rho$:}}  To sample $\rho$, we use a random walk Metropolis-Hasting step on the logit scale. More specically Let $\rho'$ denote the proposed value for $\rho$ which is obtained as
$$
\rho' = \frac{1}{1 + \exp \left\{ -\log\frac{\rho}{1-\rho} + \nu\right\}},
$$
where $\nu$ follows a normal distribution with mean 0 and variance $\tau^2$.
Then $\rho'$ is accepted with probability:
\begin{align*}
    \min \left\{ 1, \frac{\left[ \prod_{i = 1}^I\phi(\v{\beta_i} \mid \v{0}, \bfOmega(\rho')) \right] \phi_{(0, 1)}(\rho' \mid \eta, \lambda^2) \rho' (1 - \rho')}
    {\left[ \prod_{i = 1}^I\phi(\v{\beta_i} \mid \v{0}, \bfOmega(\rho)) \right] \phi_{(0, 1)}(\rho \mid \eta, \lambda^2) \rho (1 - \rho)} \right\}.
\end{align*}
where $\phi( \cdot \mid \bfmu, \bfSigma)$ denotes the density of a (multivariate) normal distribution with mean $\bfmu$ and variance $\bfSigma$, and $\phi_{(0, 1)}( \cdot \mid \eta, \lambda^2)$ denotes the density of a (univariate) normal distribution with mean $\eta$ and variance $\lambda^2$, truncated to the interval $[0,1]$.

\begin{figure}[!t]
    \centering
    \begin{subfigure}[t]{.48\textwidth}
    \centering
    \includegraphics[width = 0.95\textwidth]{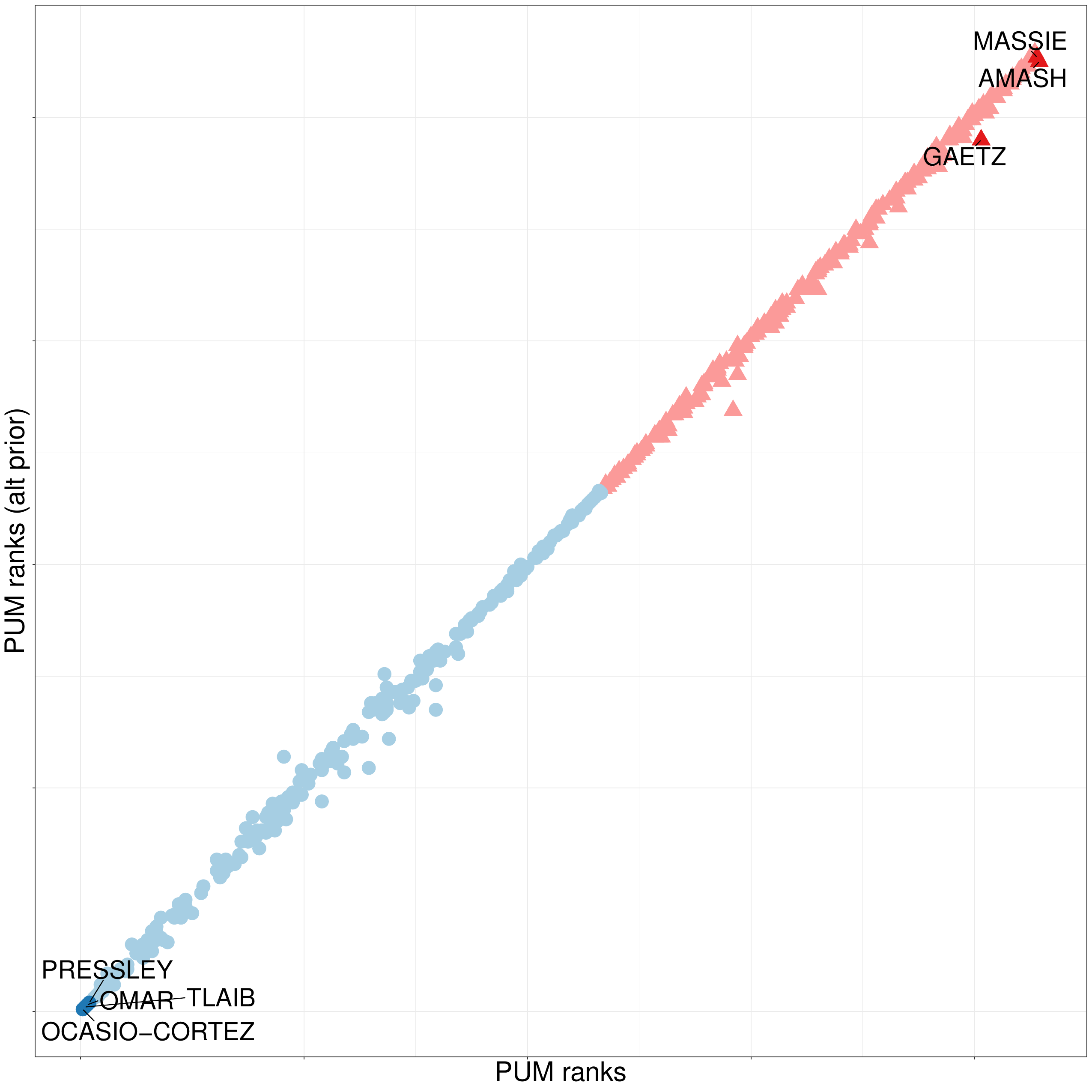}
    \caption{Posterior median ranks}
    \end{subfigure}
    \begin{subfigure}[t]{.48\textwidth}
    \centering
    \includegraphics[width =0.95\textwidth]{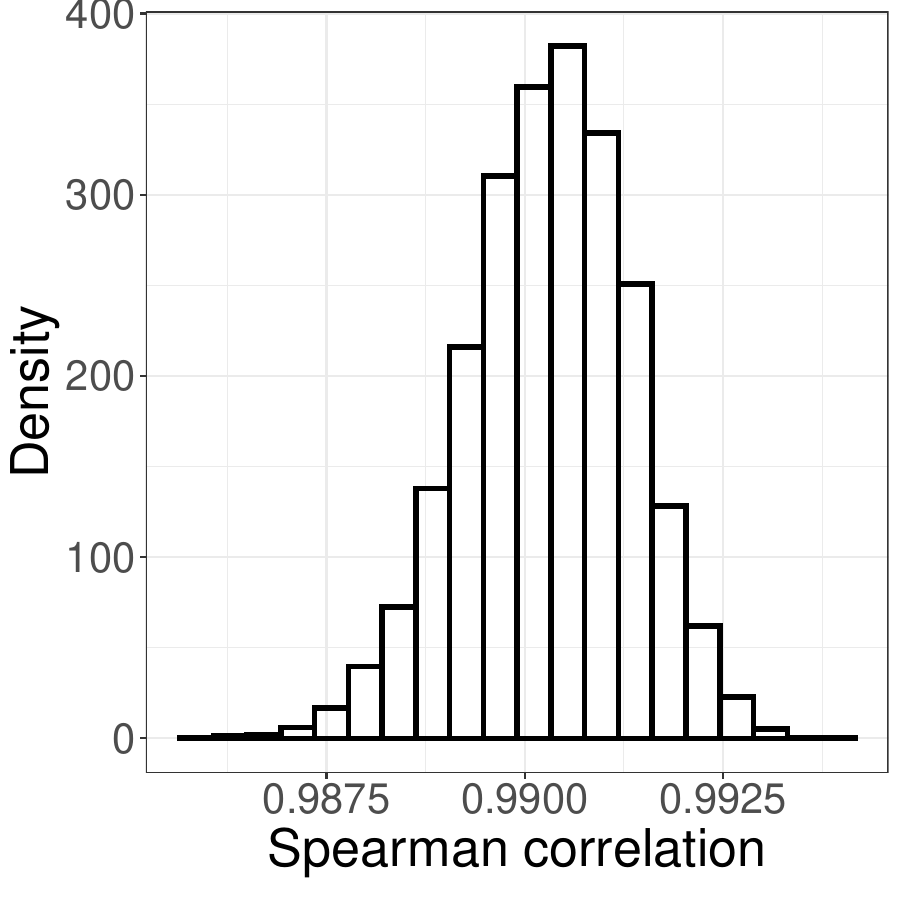}
    \caption{Spearman rank correlation}
    \end{subfigure}
    \caption{Comparison of the posterior distribution over legislator's ranks for the 116\textsuperscript{th} House under two alternative priors.  The left panel shows a comparison of the median ranks for each legislator. Democrats are shown with blue triangles, Republicans are shown with red triangles, and independents are shown with a rhombus and the color of the party that they caucus with.  The right panel shows the posterior distribution of the Spearman correlation between both sets of ranks.}
    \label{fig:leg_ranks_prior_comp}
\end{figure}

\section{Evolution of inter-party spread in revealed preferences}

\begin{figure}[!t]
\centering
    \centering
    \includegraphics[width = 0.7\textwidth]{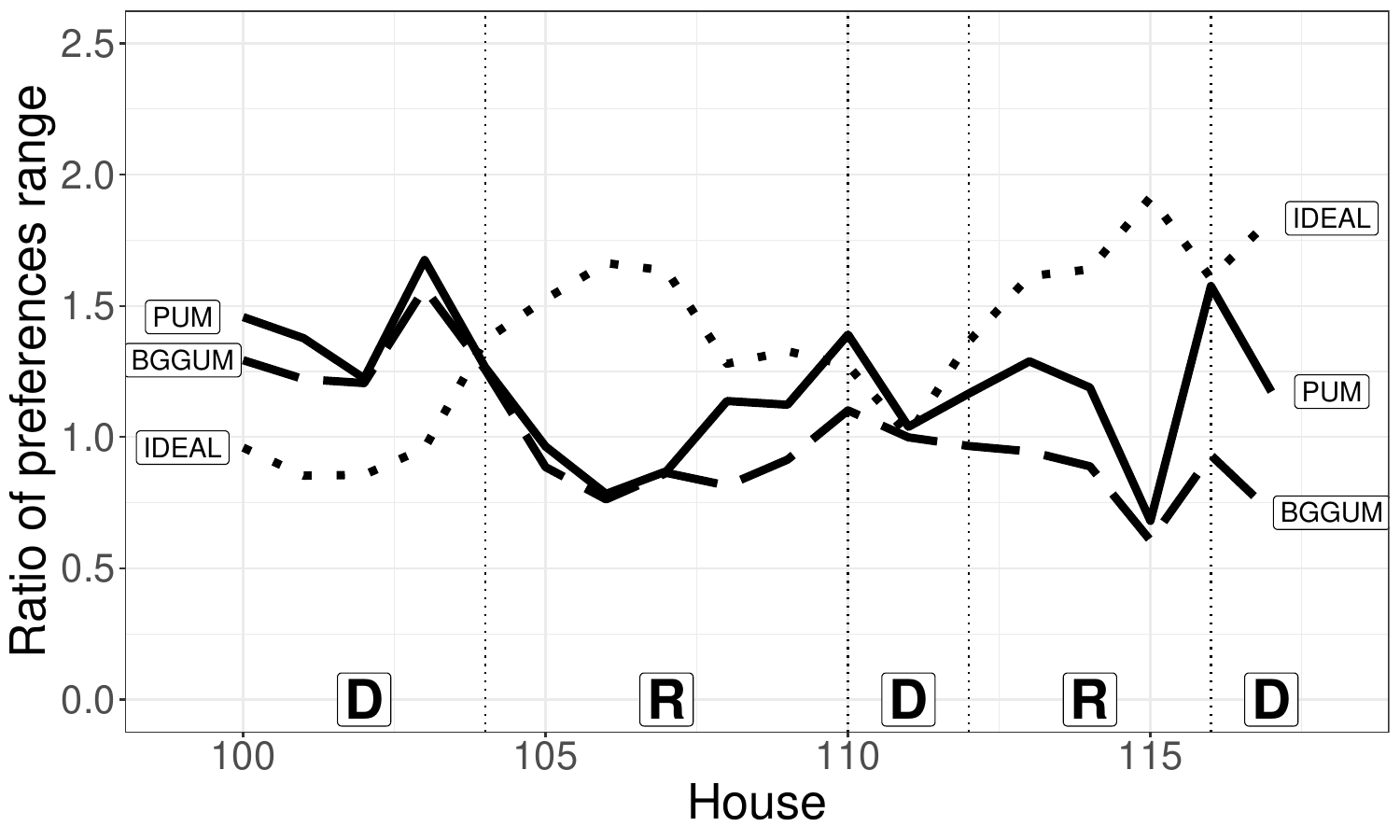}
    \includegraphics[width = 0.7\textwidth]{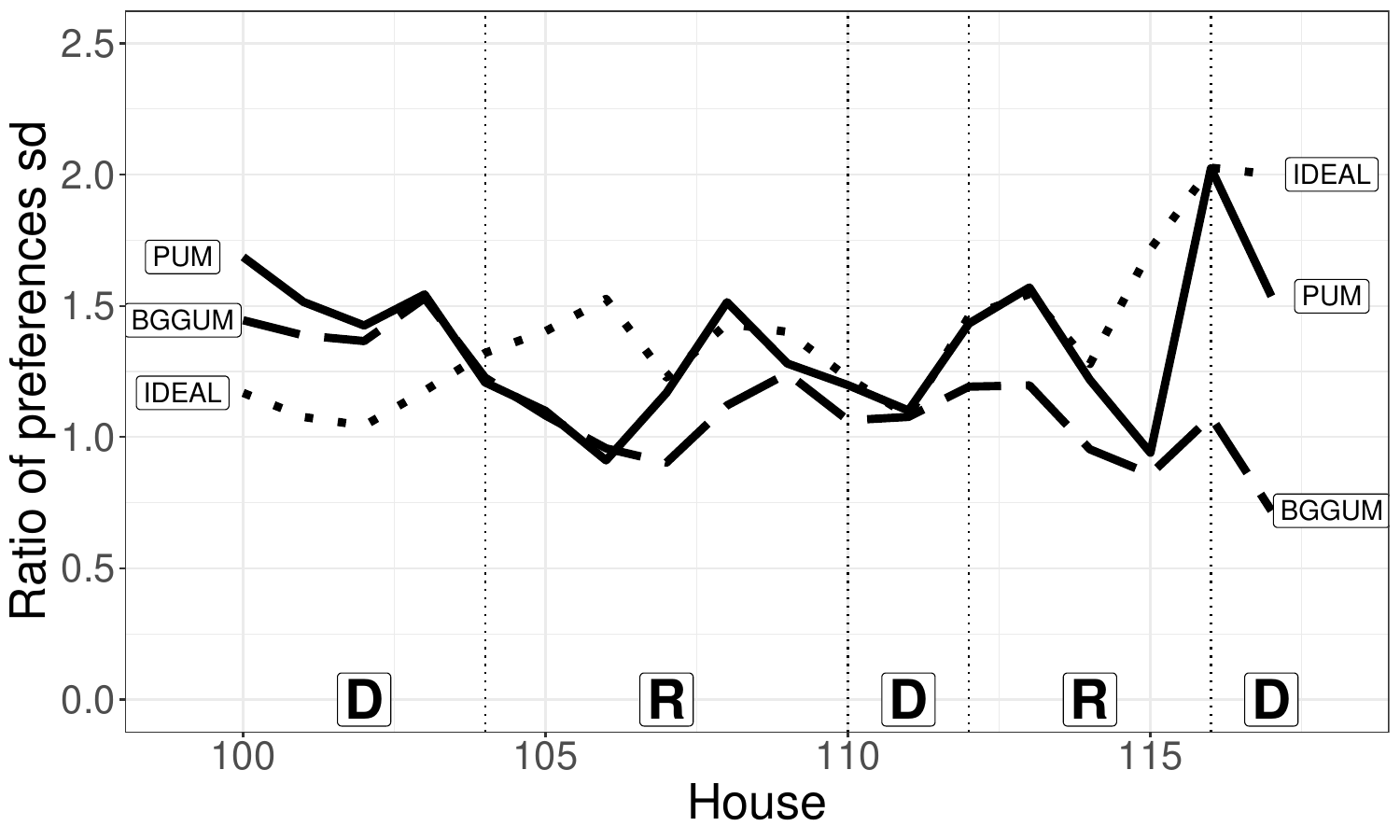}
    \includegraphics[width = 0.7\textwidth]{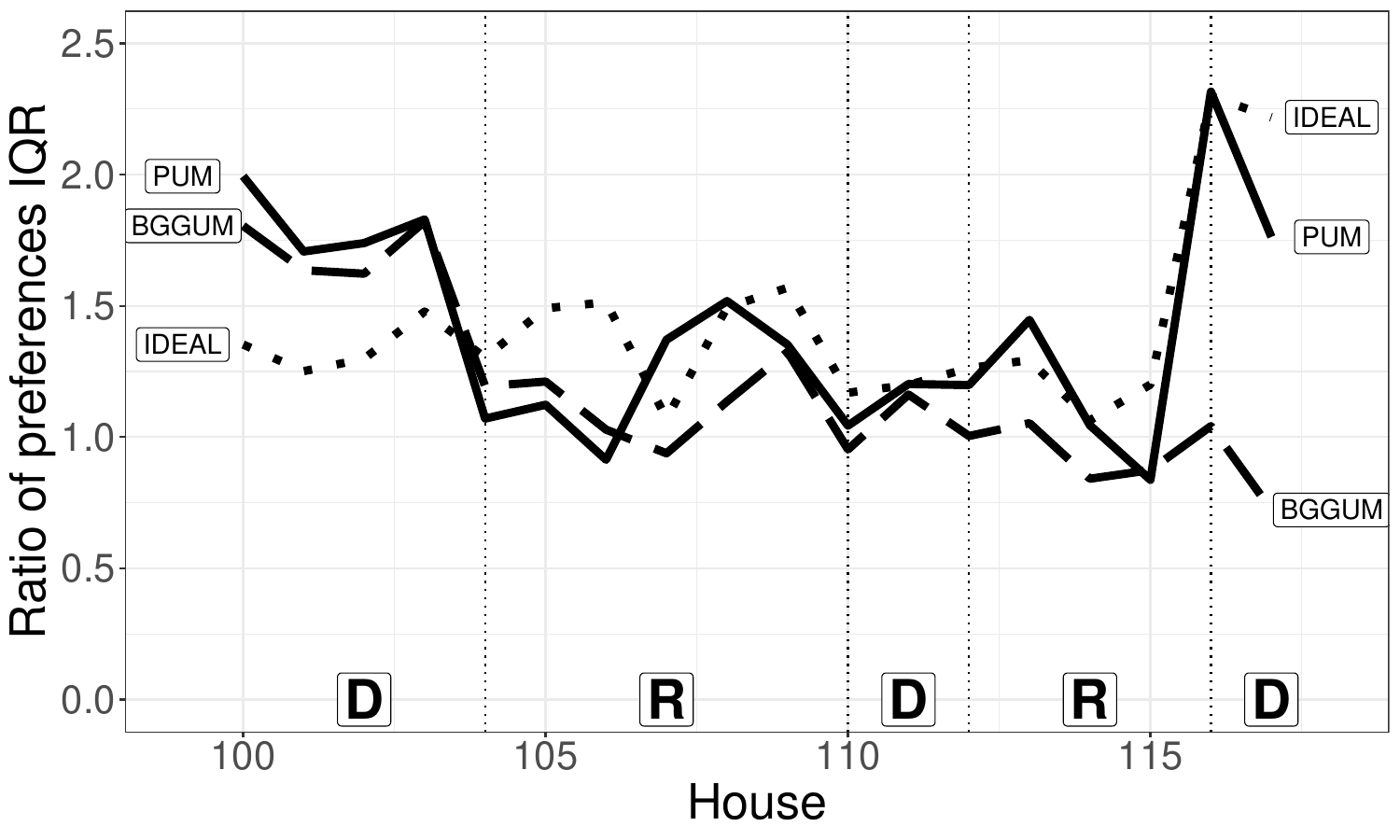}
\caption{Posterior mean of the ratio of Democrats' range, standard deviation and interquartile range (IQR) over the same metric for Republicans across the various Houses. The solid line corresponds to the probit unfolding model, the dotted line to IDEAL, and the dashed line to BGGUM.}
\label{fig:static_ratio_comp}
\end{figure}

Figure \ref{fig:static_ratio_comp} presents the ratio of the spread of Democrats' revealed preferences to the spread of Republican's preferences based on three alternative metrics of spread:  range, standard deviation and interquartile range (IQR).  The first plot just repeats Figure 5 of the main manuscript.  We can see that the overall picture is very similar no matter what metric is used, but that there some minor differences.  For example, the differences between IDEAL and the probit unfolding model that we observed for the range between the 112\textsuperscript{th} and 115\textsuperscript{th} House are less pronounced under the standard deviation, and even less so for the interquartile range.

\section{Sensitivity analysis for the static probit unfolding model}
\label{sec:static_sensitivity_analysis}

In this section, we conduct a sensitivity analysis for the results associated with the 116\textsuperscript{th} U.S.\ House of Representatives (see Section 3 of the main manuscript).  To accomplish this, we refit the model setting $\omega^2 =  1$ and $\kappa^2 = 9$.  The implied prior on $\theta_{i,j} = \Pr\left(y_{i,j} = 1 \mid \beta_i, \bfalpha_j, \bfdelta_j \right)$ can be seen on the left panel of Figure \ref{fig:prior_draws_prior_comp}. To facilitate comparison, the right panel shows again the prior on $\theta_{i,j}$ implied by the original prior (in which $\omega^2 =  25$ and $\kappa^2 = 10$, recall Figure 2a in the main manuscript).  Note that the alternative prior favors values of $\theta_{i,j}$ close to 0.5 much more strongly than the original prior.

\begin{figure}[!t]
\centering
\begin{subfigure}[t]{.3\textwidth}
    \centering
    \includegraphics[width = .95\textwidth]{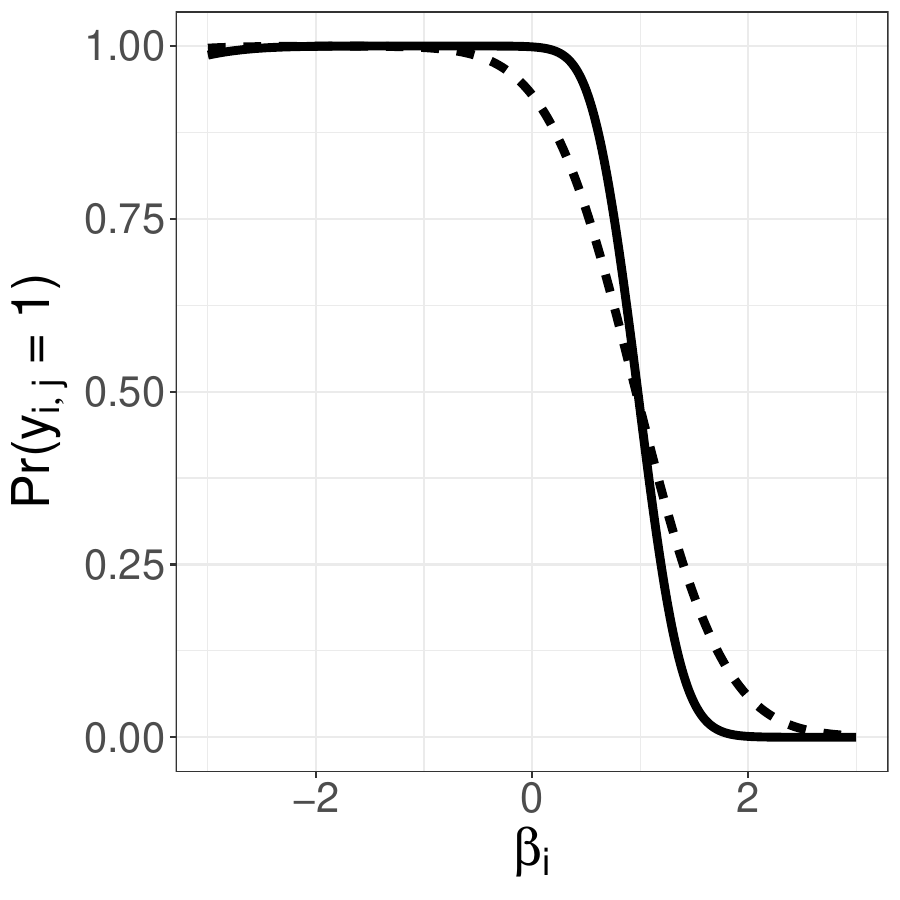}
    \caption{HRES5}
\end{subfigure}
\begin{subfigure}[t]{.3\textwidth}
    \centering
    \includegraphics[width = .95\textwidth]{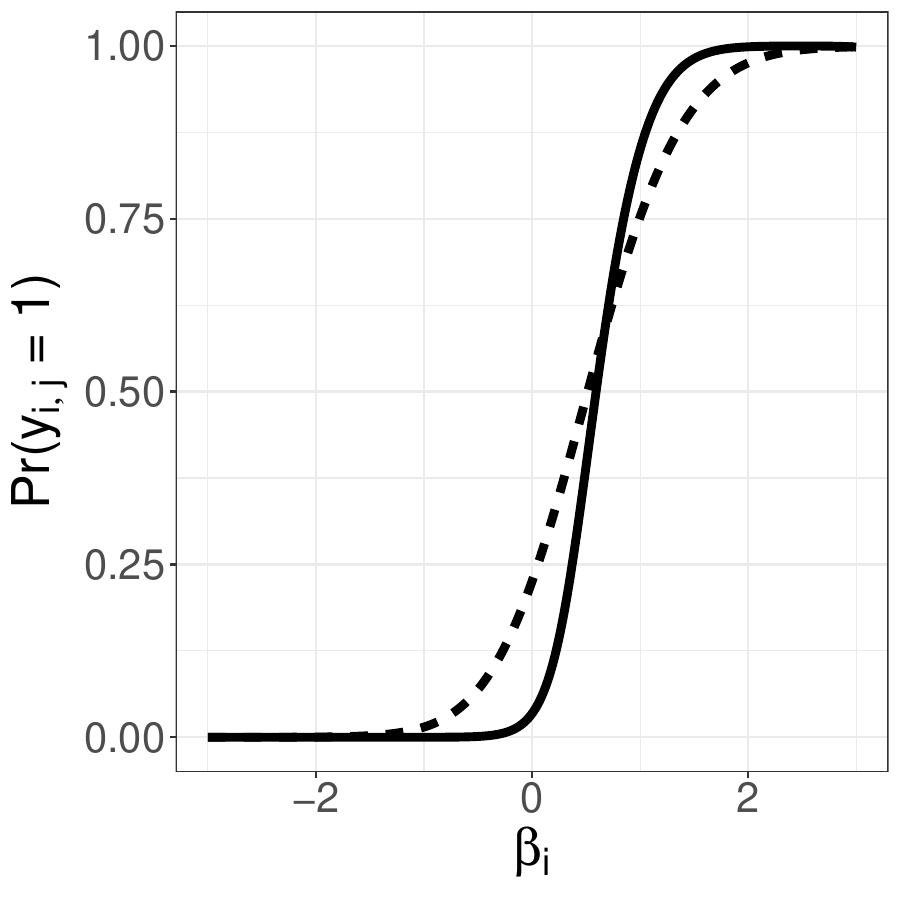}
    \caption{HR21}
\end{subfigure}
\begin{subfigure}[t]{.3\textwidth}
    \centering
    \includegraphics[width = .95\textwidth]{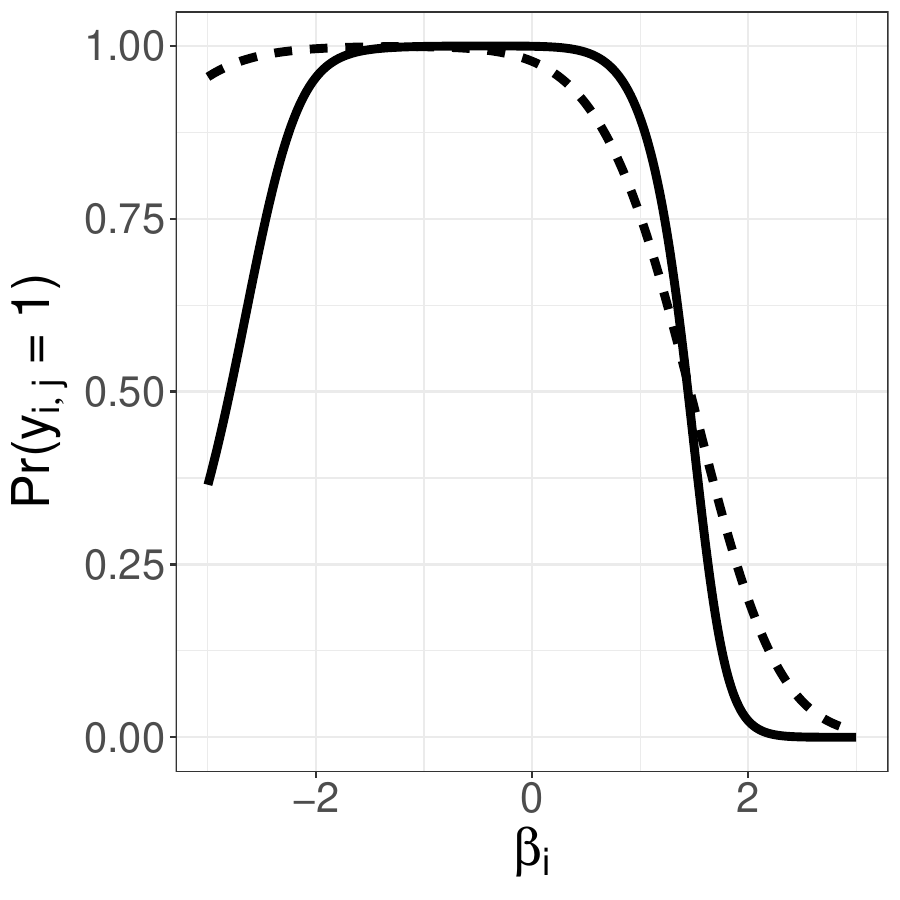}
    \caption{HRES6}
\end{subfigure}
\caption{Plots displaying various posterior mean response curves based on the posterior means of $\alpha_{j,1}, \alpha_{j,2}$, $\delta_{j,1}$, and $\delta_{j,2}$ from the probit unfolding model for the 116\textsuperscript{th} House under the two priors. The response curves for the prior discussed in the paper is shown with a solid line whereas the response curves for the alternative prior is shown with dashed lines.}
\label{fig:response_curves_116_prior_comp}
\end{figure}

The left panel of Figure \ref{fig:leg_ranks_prior_comp} compares the posterior median ranks of legislators under both priors, while its right panel shows the posterior distribution of their Spearman correlation. These graphs suggest that the ranks are fairly robust to the change in prior parameters.  On the other hand, Figure \ref{fig:response_curves_116_prior_comp} compares the posterior mean response curves associated with the same three votes discussed in Figure 7 of the main manuscript. Here we see a more pronounced effect of the prior, specially for HRES6.  These differences, which mainly seem to be associated with the slopes of the response function, are not too surprising.  Indeed, the alternative prior has a much lower variance for $\alpha_{j,1}$ and $\alpha_{j,2}$ than the original, which discourages large values for the slopes.

Finally, we also explored the impact of the alternative prior distribution on the WAIC score of our model (recall Equation (5) in the main manuscript).  Interestingly, the WAIC under the alternative prior is worse than the WAIC under the original one, and also worse than the WAIC for BGGUM. However, the WAIC score for our model under the alternative prior is still better than the WAIC for IDEAL.  These results suggest that our observation that unfolding models generally seem to better explain voting patterns in the modern U.S.\ House of Representatives is fairly robust to prior choices.  However, it also reminds us that priors can have a substantial effect on model comparison criteria, and that matching priors (as we did in Section 2.1 of the main manuscript) is an important prerequisite to any meaningful application of  model selection criteria.

\section{Sensitivity analysis for dynamic probit unfolding model}
\label{sec:dynamic_sensitivity_analysis}

We now study the sentivity to prior choices in the context of the analysis of U.S.\ Supreme Court vote data discussed in Section 5 of the main manuscript. For this sensitivity analysis, we consider an alternative prior on the autocorrelation parameter $\rho$ that corresponds to a normal distribution with mean $0.8$ and standard deviation $0.1$, truncated to the $[0, 1]$ interval.
\begin{figure}
\begin{subfigure}[t]{.45\textwidth}
    \centering
    \includegraphics[width = .95\textwidth]{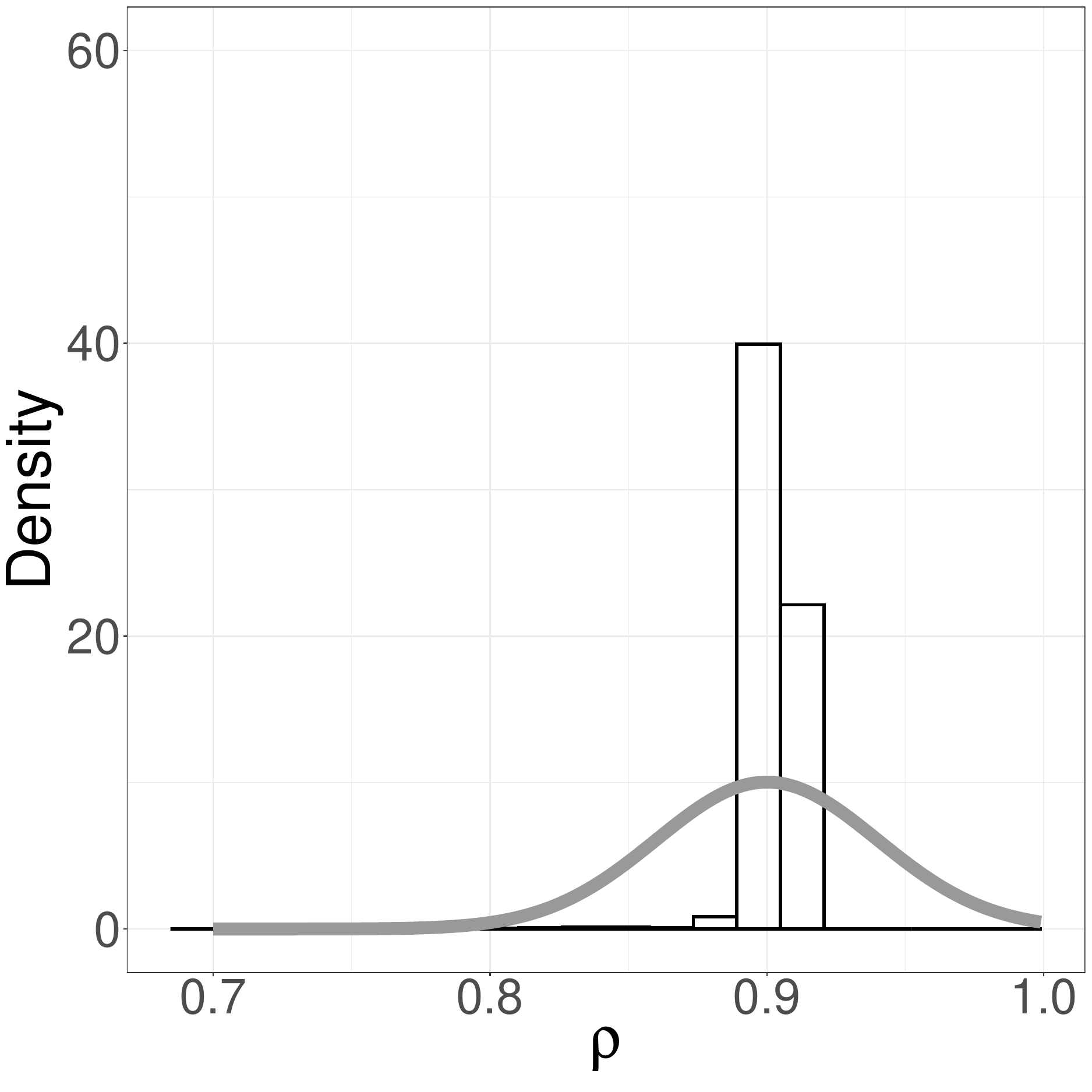}
    \caption{Original prior}
\end{subfigure}
\begin{subfigure}[t]{.45\textwidth}
    \centering
    \includegraphics[width = .95\textwidth]{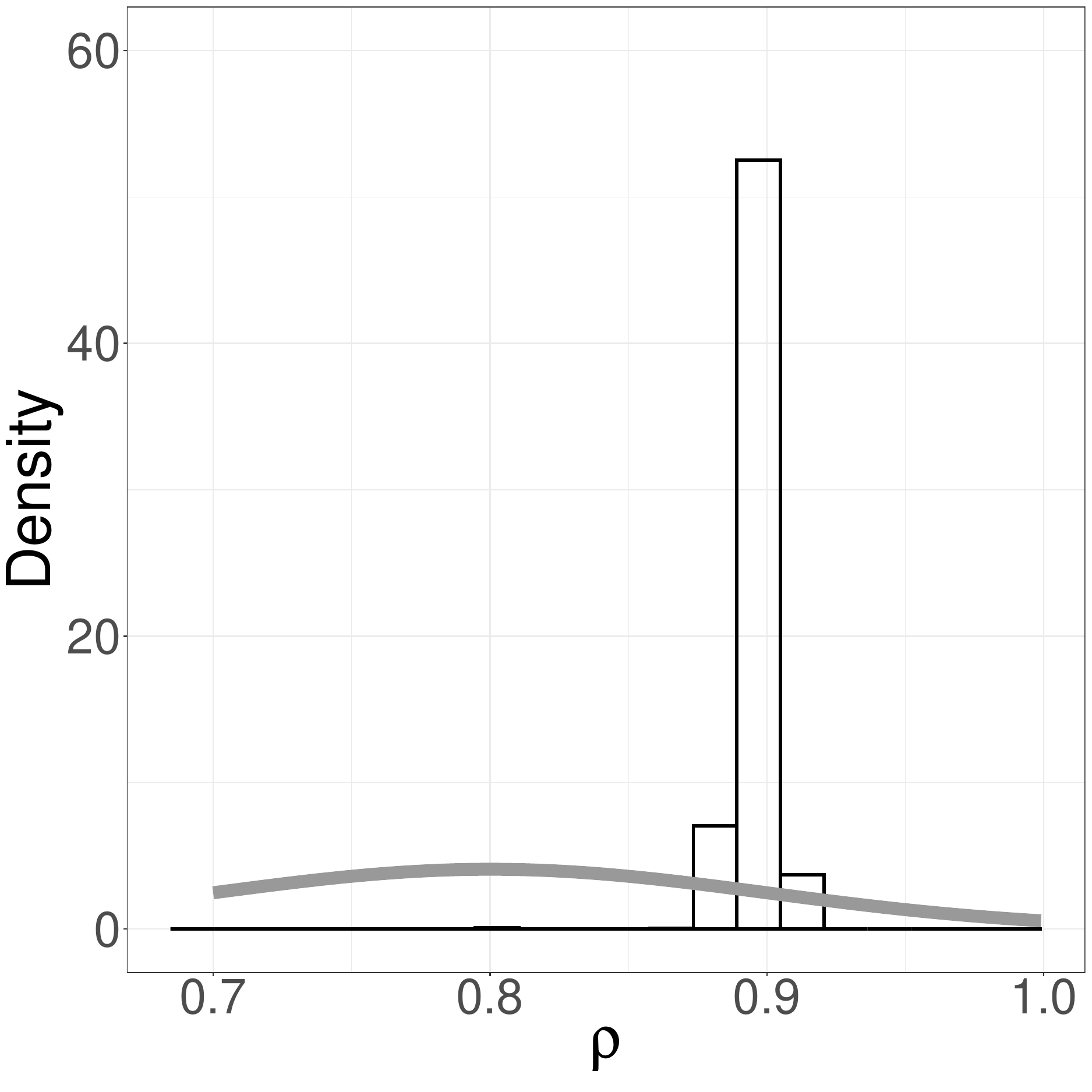}
    \caption{Alternative prior}
\end{subfigure}
\caption{Histograms for the posterior of $\rho$ under various priors with its respective prior probability density function plotted in grey.}
\label{fig:prior_draws_prior_comp_dynamic}
\end{figure}

Figure \ref{fig:prior_draws_prior_comp_dynamic} displays histograms of the posterior draws of $\rho$ against the plots of priors' probability density function in each of the two scenarios. As seen in this plot, the alternative prior is centered at a much lower value for $\rho$ and is much more diffuse. Indeed, under the original prior, $P(\rho < 0.85)$ is roughly 0.1, while under the alternative prior, $P(\rho < 0.85)$ is roughly 0.7. However, despite the fact the new prior places less probability on large values of $\rho$, the posterior is still quite concentrated around 0.9. This suggests that the posterior distribution of $\rho$ is fairly robust to the prior.
\begin{figure}[!b]
    \centering
    \begin{subfigure}[t]{0.49\textwidth}
        \centering
        \includegraphics[width = \textwidth]{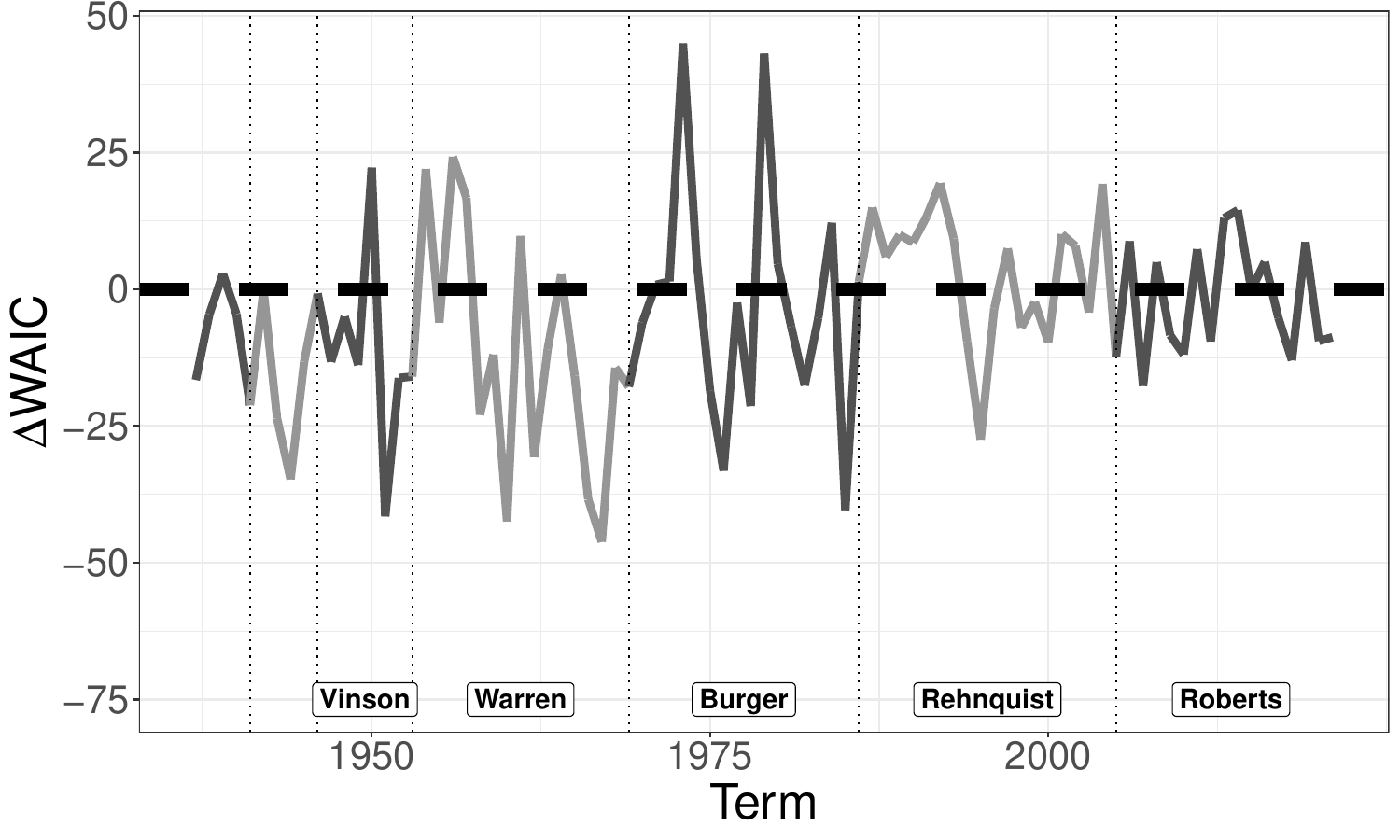}
        \caption{WAIC}\label{fig:WAIC_comparison_SCOTUS_a}
    \end{subfigure}
    \begin{subfigure}[t]{0.49\textwidth}
        \centering
        \includegraphics[width = \textwidth]{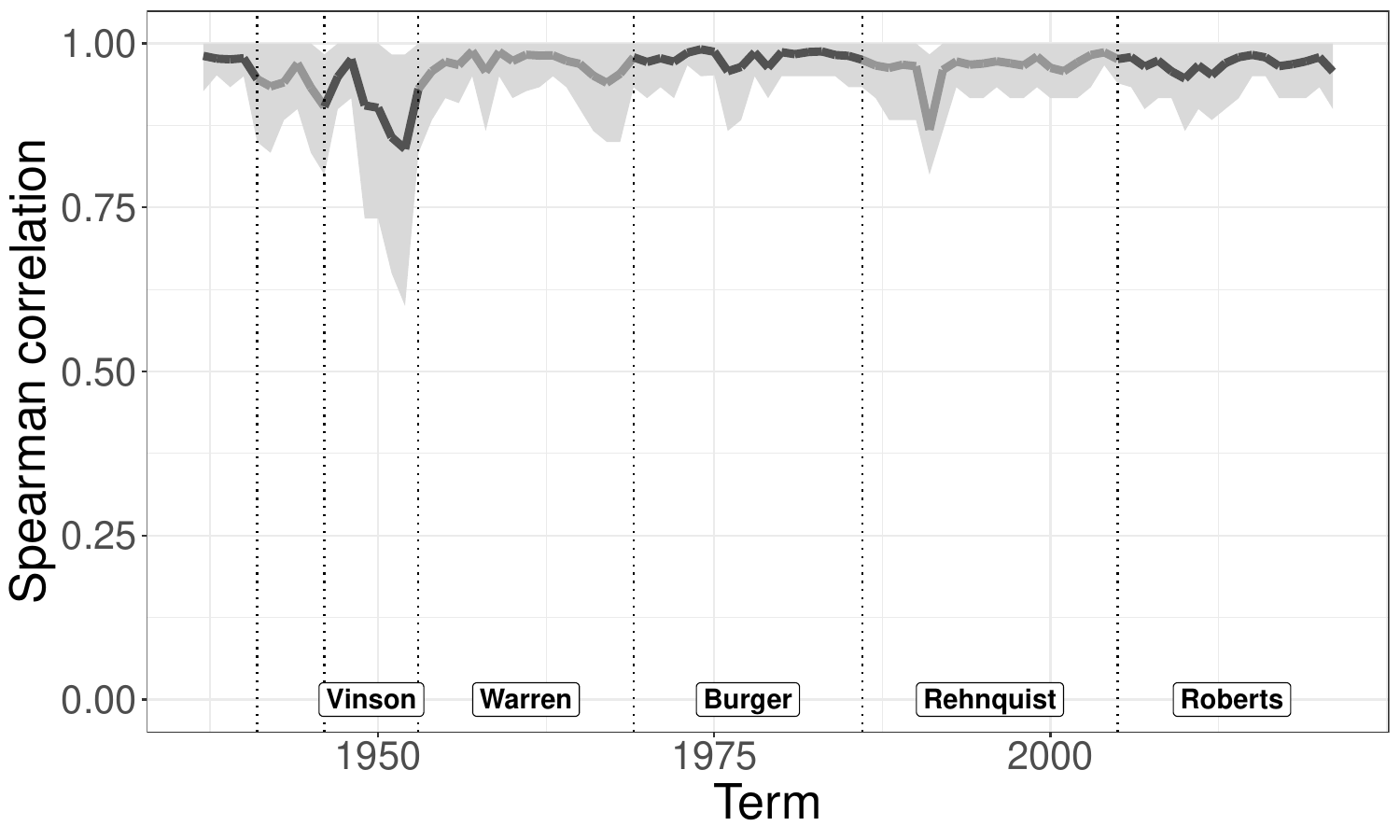}
        \caption{Spearman correlation}\label{fig:rank_comparison_SCOTUS_a}
    \end{subfigure}
\caption{Left panel: Difference in WAIC scores for dynamic unfolding model under the original and alternative prior on $\rho$, ($WAIC$(DPUM alternative prior) - $WAIC$(DPUM original prior)). Note that the way the difference is being computed here is the opposite to the way in which it was computed in Figure 3a in the main text. Right panel: Posterior mean (solid line) and corresponding 95\% credible intervals (shaded region) for the Spearman correlation between the justices' rankings generated under the two priors for $\rho$.}
\end{figure}

\begin{figure}[!t]
\centering
\begin{subfigure}[t]{0.45\textwidth}
    \centering
    \includegraphics[width = .95\textwidth]{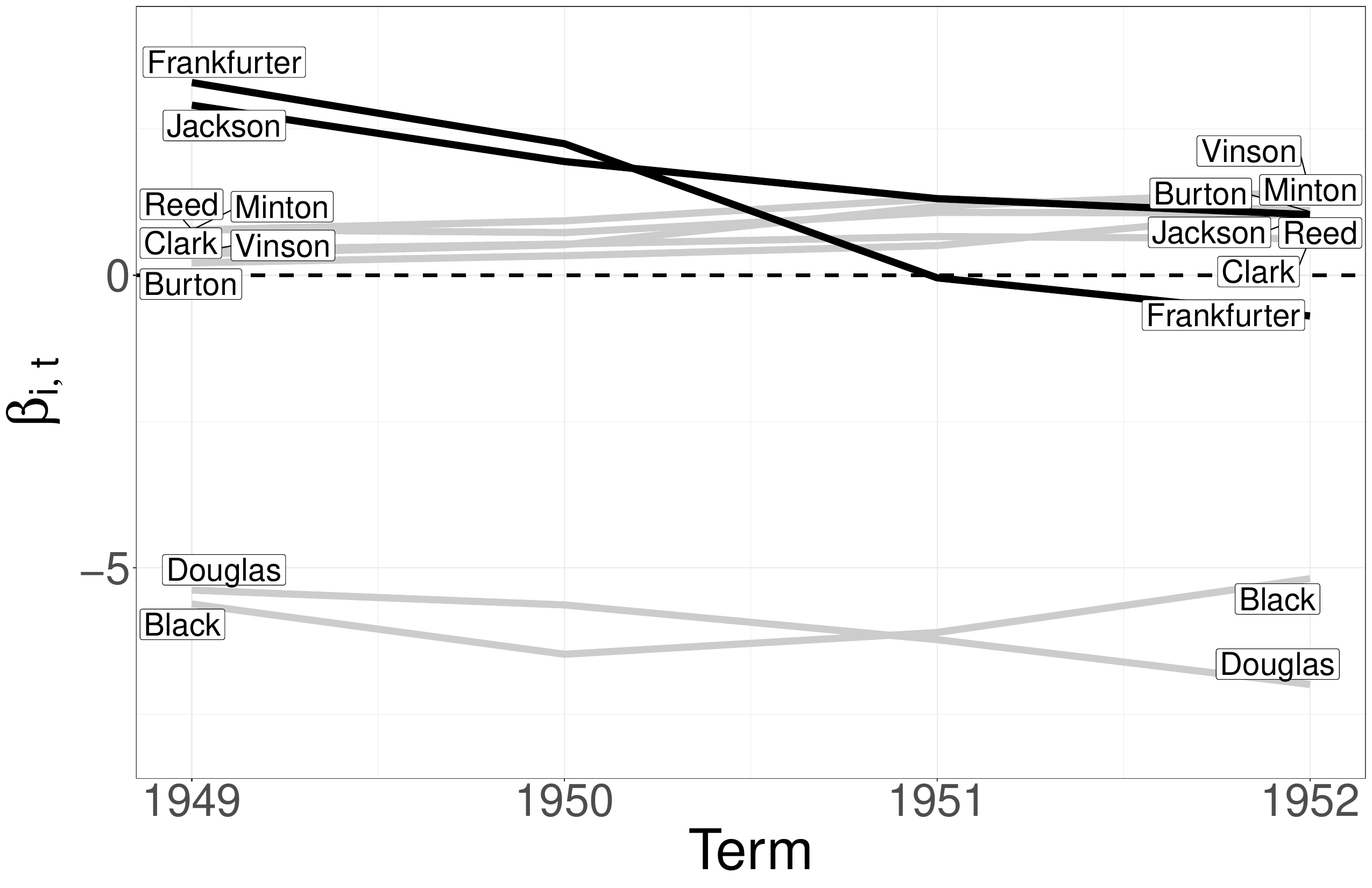}
    \caption{Original prior}\label{fig:scores1949to1952threeup_a}
\end{subfigure}
\begin{subfigure}[t]{0.45\textwidth}
    \centering
    \includegraphics[width = .95\textwidth]{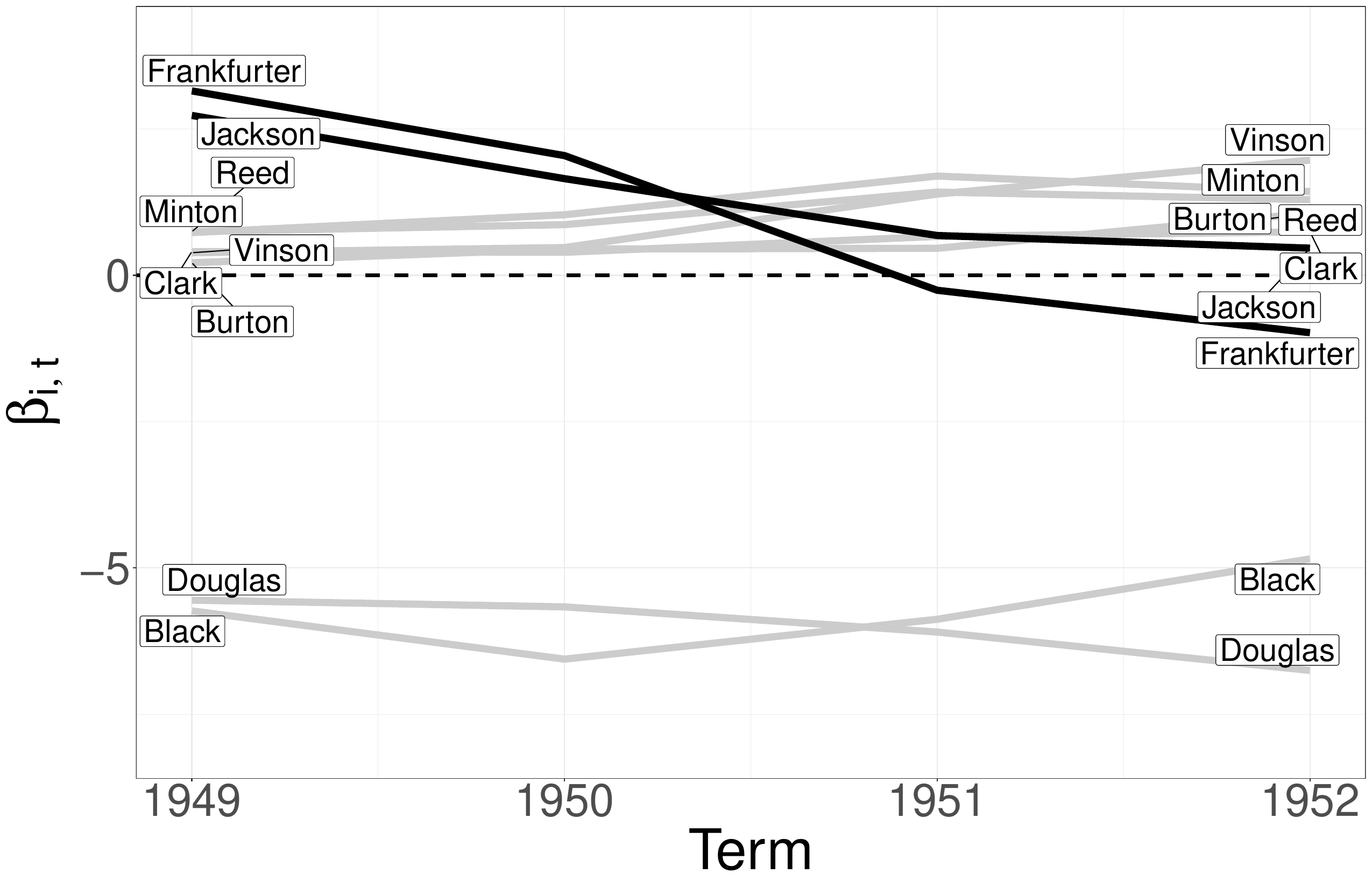}
    \caption{Alternative prior}\label{fig:scores1949to1952alt}
\end{subfigure} 
\caption{Posterior mean of the ideal points for SCOTUS Justices active during 1949 to 1952 terms under the dynamic unfolding model for various priors for $\rho$.
} \label{fig:scores1949to1952_alt}
\end{figure}

We now examine the downstream effect of this prior choice. Figure \ref{fig:rank_comparison_SCOTUS_a} shows the posterior mean (solid line) and corresponding 95\% credible intervals (shaded region) for the Spearman correlation between the justices' rankings generated under the two priors we consider. Overall, we can see that the ranks are quite similar, although there seems to be some sensitivity during the 1949-1952 term, and perhaps again in the 1991 term.  On the other hand, Figure \ref{fig:WAIC_comparison_SCOTUS_a} shows the difference in WAIC scores between the original and  alternative prior on $\rho$.  This graph does not show a clear pattern in favor of either prior.

To better understand the sensitivity of the ranks to the prior, we present in Figure \ref{fig:scores1949to1952_alt} the posterior mean of the ideal points of SCOTUS justices active during 1949 to 1952 under both the original and the alternative prior. The main difference between the two plots relates to the trajectories of Justices Felix Frankfurter and Robert Jackson.  Under the alternative prior (which favors lower values of $\rho$ and, therefore, allows for less consistent preferences over time) the ideal points of Justices Frankfurter and Jackson drop slightly faster in 1951 and 1952 than under the original prior.  This, together with the fact that Justices Reed, Burton, Vinson, Clark and Minton tend to vote together fairly often (leading to very similar estimates of their ideal points) means that even a small change in the estimate of Justice Frankfurter and Jackson's ideal points can have have a large impact on the associated ranks.